\documentclass[amsmath,trackchanges, twocolumn]{aastex702} % linenumbers
\usepackage{graphicx}
\begin{document}

\title{DETECT: Real-Time Identification of Transients with DESI Spectroscopic Redshift}

% ===============================
%  Corresponding authors
% ===============================
\correspondingauthor{Y.-H.~Lee}
\email[show]{m1139005@astro.ncu.edu.tw}
\correspondingauthor{T.-W.~Chen}
\email[show]{twchen@astro.ncu.edu.tw}

% ===============================
%  Author list
% ===============================
\author[orcid=0009-0003-5139-9007,sname=Lee,gname=Yu-Hsing]{Yu-Hsing Lee}
\affiliation{Graduate Institute of Astronomy, National Central University, 300 Jhongda Road, 32001 Jhongli, Taiwan}
\email[]{m1139005@astro.ncu.edu.tw}

\author[0000-0002-1066-6098,sname=Chen,gname=Ting-Wan]{Ting‑Wan Chen}
\affiliation{Graduate Institute of Astronomy, National Central University, 300 Jhongda Road, 32001 Jhongli, Taiwan}
\email[]{twchen@astro.ncu.edu.tw}

\author[sname=Wang,gname=Ze-Ning]{Ze-Ning Wang}
\affiliation{Henan Academy of Sciences, Zhengzhou 450046, Henan, China}
\affiliation{Institute of Astrophysics, Central China Normal University, Wuhan 430079, China}
\email[]{}

\author[0000-0002-2898-6532,sname=Yang,gname=Sheng]{Sheng Yang}
\affiliation{Institute for Gravitational Wave Astronomy, Henan Academy of Sciences, Zhengzhou 450046, Henan, China}
\email[]{sheng.yang@hnas.ac.cn}

\author[0009-0009-9255-920X,sname=Deng,gname=Limeng]{Limeng Deng}
\affiliation{Technical University of Munich, TUM School of Natural Sciences, Physics Department, James-Franck-Str. 1, 85748 Garching, Germany}
\affiliation{Max Planck Institute for Astrophysics, Karl-Schwarzschild-Str. 1, 85748 Garching, Germany}
\email[]{}

\author[0000-0003-1449-7284,sname=Chuan-Jui,gname=Li]{Chuan-Jui Li}
\affiliation{Graduate Institute of Applied Physics, National Chengchi University, Taipei 116026, Taiwan}
\email[]{}

% ======================================
%PS
\author[0000-0001-6965-7789,sname=Chambers,gname=K.~C.]{K.~C. Chambers}
\affiliation{Institute for Astronomy, University of Hawai`i, 2680 Woodlawn Drive, Honolulu, HI 96822, USA}
\email[]{}

\author[0000-0001-5486-2747,sname=de~Boer,gname=Thomas]{Thomas de Boer}
\affiliation{Institute for Astronomy, University of Hawai`i, 2680 Woodlawn Drive, Honolulu, HI 96822, USA}
\email[]{}

\author[0000-0002-7272-5129,sname=Lin,gname=Chien-Cheng]{Chien-Cheng Lin}
\affiliation{Institute for Astronomy, University of Hawai`i, 2680 Woodlawn Drive, Honolulu, HI 96822, USA}
\email[]{}

\author[0000-0002-9438-3617,sname=Lowe,gname=Thomas~B.]{Thomas B. Lowe}
\affiliation{Institute for Astronomy, University of Hawai`i, 2680 Woodlawn Drive, Honolulu, HI 96822, USA}
\email[]{}

\author[0009-0003-8803-8643,sname=M\'inguez,gname=Paloma]{Paloma M\'inguez}
\affiliation{Institute for Astronomy, University of Hawai`i, 2680 Woodlawn Drive, Honolulu, HI 96822, USA}
\email[]{}

\author[0000-0002-6639-6533,sname=Paek,gname=Gregory~S.~H.]{Gregory S.~H. Paek}
\affiliation{Institute for Astronomy, University of Hawai`i, 2680 Woodlawn Drive, Honolulu, HI 96822, USA}
\email[]{}

\author[0000-0002-1341-0952,sname=Wainscoat,gname=Richard]{Richard Wainscoat}
\affiliation{Institute for Astronomy, University of Hawai`i, 2680 Woodlawn Drive, Honolulu, HI 96822, USA}
\email[]{}

% ===============================
% Abstract
% ===============================
\begin{abstract}
Wide-field surveys now report thousands of transients per month, while spectra are obtained for only a few per cent of them. We present DETECT (DESI--Transient Event Cross-matching Tool), a pipeline that turns each Transient Name Server (TNS) alert into a distance-informed candidate within an hour by placing it in the context of the Dark Energy Spectroscopic Instrument (DESI) archive. DETECT cross-matches every new report against the 22 million spectra of DESI EDR and DR1 through a HEALPix index, associates the transient with a host galaxy by a directional-light-radius rule whose morphology-dependent thresholds are calibrated on 4,474 supernovae with known redshifts (92.5\% completeness, 0.4\% wrong hosts), and passes ambiguous cases to a reviewer through a web interface. With a validated host redshift it derives the peak absolute magnitude, the projected offset and basic host properties of each event, and ranks it for spectroscopic follow-up. Applied retrospectively to $\sim1.2\times10^5$ TNS reports from 2020--2024, DETECT recovers a spectroscopic host for 15\% of them, from these we build a Gold Sample of $\sim$5,400 well-sampled light curves whose luminosity distributions by class reproduce those of untargeted surveys. Because the absolute magnitude is known at discovery, intrinsically overluminous events stand out immediately: in the prospective 2025 run this is how the lensed superluminous supernova SN~2025wny was flagged within hours of its report, and how faint, fast-declining kilonova candidates were screened against model grids and archival photometry. DETECT shows how an archival spectroscopic survey can make host redshifts a routine part of transient triage ahead of the Rubin Observatory's LSST.
\end{abstract}

% ===============================
% Keywords
% ===============================
% =============================================================
% MAIN TEXT
% =============================================================
\section{Introduction}
\label{sec:intro}
Wide-field time-domain surveys such as ATLAS \citep{Tonry_2018}, Pan-STARRS1 \citep{2016arXiv161205560C}, ZTF \citep{2019PASP..131a8002B, 2019PASP..131g8001G}, GOTO \citep{2022MNRAS.511.2405S} and BlackGEM \citep{2024PASP..136k5003G} have substantially increased the discovery rate of astrophysical transients. Alert volumes are expected to grow further in the Rubin era \citep[e.g.][]{2022ApJS..258....1B}. However, spectroscopic follow-up resources remain limited: dedicated classification programmes such as PESSTO/ePESSTO+ \citep{2015A&A...579A..40S} and the ZTF Bright Transient Survey \citep{2020ApJ...895...32F} classify a few thousand events per year, essentially the bright end of the population, making it impractical to obtain spectra for more than a small fraction of newly reported events. As a result, photometric and contextual methods have become increasingly important for transient classification, follow-up selection, and population studies. Alert brokers such as ALeRCE \citep{2021AJ....161..242F}, Fink \citep{2021MNRAS.501.3272M}, Lasair \citep{2019RNAAS...3...26S}, ANTARES \citep{2021AJ....161..107M} and AMPEL \citep{2019A&A...631A.147N} annotate the ZTF stream with light-curve classifiers and catalogue cross-matches, but a spectroscopic host redshift, and hence an absolute magnitude at discovery, is available for only a small fraction of their alerts.

A host redshift enables an estimate of the distance modulus and thus the transient absolute magnitude, which can be more informative than apparent magnitude alone for identifying intrinsically luminous or otherwise unusual events. This is particularly valuable for rare and extreme transient populations. Superluminous supernovae \citep[SLSNe,][]{2011Natur.474..487Q,2019ARA&A..57..305G}, for example, are generally distinguished by their high peak luminosities and their preference for relatively faint dwarf host galaxies \citep{2011ApJ...727...15N, 2013ApJ...763L..28C, 2014ApJ...787..138L, 2015MNRAS.449..917L, 2016ApJ...830...13P, 2017ApJ...849L...4C, 2018MNRAS.473.1258S}. In practice, a transient with an inferred peak magnitude of roughly $M \lesssim -21$ mag \citep{2012Sci...337..927G} may be considered a plausible SLSN candidate, although the exact physical and phenomenological boundary between SLSNe and the luminous tail of more ordinary SN populations remains unsettled \citep[e.g.][]{2021MNRAS.508.4342P, 2016A&A...596A..67R}. Similarly, a transient whose inferred luminosity is far too high for a normal supernova may be a background event magnified by a foreground lens \citep[e.g.][]{2025arXiv250916033H, 2025arXiv251204275K}, a host redshift is what makes such an excess visible at discovery, before any lens-specific search is made.

Distance information is also important at the faint end of the transient luminosity distribution. In particular, kilonovae are intrinsically less luminous than most SN populations and fade on very short timescales \citep[e.g.][]{2017Natur.551...75S,2017ApJ...848L..27T}, making them difficult to detect in survey data with insufficient cadence \citep[e.g.][]{2021MNRAS.500.4213M,2021ApJ...918...63A,2025MNRAS.542..541F}.
More broadly, approximate distance information for large transient samples enables population-level studies, including luminosity functions, class-dependent luminosity distributions \citep[e.g.][]{2020ApJ...895...32F, 2026MNRAS.549g1028S}, and the systematic identification of outliers for rapid follow-up.

Large spectroscopic galaxy surveys offer an important source of physical context for newly discovered transients \citep[e.g.][]{2025ApJ...992..158F}. In particular, the Dark Energy Spectroscopic Instrument (DESI) is providing redshifts for millions of extragalactic sources at unprecedented scale \citep{2022AJ....164..207D, 2023ApJ...943...68L}. DESI is closely linked to the Legacy Surveys imaging program, whose data are useful for host-galaxy association and inspection \citep{2019AJ....157..168D, 2017PASP..129f4101Z}. The utility of this combination was recently demonstrated by the MOST Hosts survey \citep{2024ApJS..275...22S}, which obtained host-galaxy spectra for a large archival transient sample. However, complementary efforts are still needed to use existing DESI and Legacy Surveys data in a systematic way for rapid transient characterization and follow-up prioritization.

To make systematic use of these resources, we have developed \textbf{DETECT} (DESI-Transient Event Cross-matching Tool), an automated pipeline that cross-matches newly reported transients with DESI spectroscopic galaxies on a hourly basis. DETECT ingests new events from the Transient Name Server (TNS)\footnote{\url{https://www.wis-tns.org/}}
, searches for nearby galaxy counterparts in the DESI spectroscopic catalog, computes basic quantities such as angular separation and transient absolute magnitude for an initial host assignment, and retrieves Legacy Surveys image cutouts for further inspection. The goal of DETECT is not to replace spectroscopic classification, but to provide a scalable contextual framework for rapid transient screening, follow-up prioritization, and population-level analyses. In addition, the pipeline flags events whose inferred luminosity is unusual for their host and redshift. Gravitationally lensed supernovae enter this way, it finds transients that are too bright for their apparent host, and a reviewer then asks whether a lens is responsible.

This framework is useful both for population-level studies of large transient samples and for the identification of unusual individual events. By incorporating host-redshift information, it enables the construction of a distance-informed transient sample for statistical analyses and improved photometric classification, while also helping to identify transients with unusual inferred luminosities. In this paper, we present the DETECT pipeline, describe the construction of a DESI-matched transient sample, and examine how host-redshift information can aid both transient classification and the discovery of rare events. A methodological contribution is a host-association rule based on the directional light radius, calibrated on the visually vetted 2020--2024 sample, which replaces nearest-neighbour matching in the real-time pipeline.

This paper is organized as follows. In Section\,\ref{sec:pipeline}, we describe the DETECT pipeline and its real-time architecture. In Section\,\ref{sec:analysis}, we present the retrospective analysis methodology and the DESI-based host association procedure. In Section\,\ref{sec:results}, we analyze the resulting luminosity distributions and rare luminous candidates. Further discussions are presented in Section\,\ref{sec:discussions}. 
Throughout this paper, all magnitudes are reported in the AB system \citep{1983ApJ...266..713O}, unless otherwise noted. For the calculation of luminosity distances and absolute magnitudes, we adopt a flat $\Lambda$CDM cosmology with parameters $H_0 = 67.4\,\mathrm{km}\,\mathrm{s}^{-1}\,\mathrm{Mpc}^{-1}$, $\Omega_{\mathrm{m}} = 0.315$, and $\Omega_{\Lambda} = 0.685$, consistent with the results from the \citet{2020A&A...641A...6P}.

% =============================================================
% =============================================================
% =============================================================

\section{DESI Cross-Match Pipeline}
\label{sec:pipeline}

\subsection{Automated Pipeline Architecture}
\begin{figure*}[tp]
    \centering
    \includegraphics[width=1\linewidth]{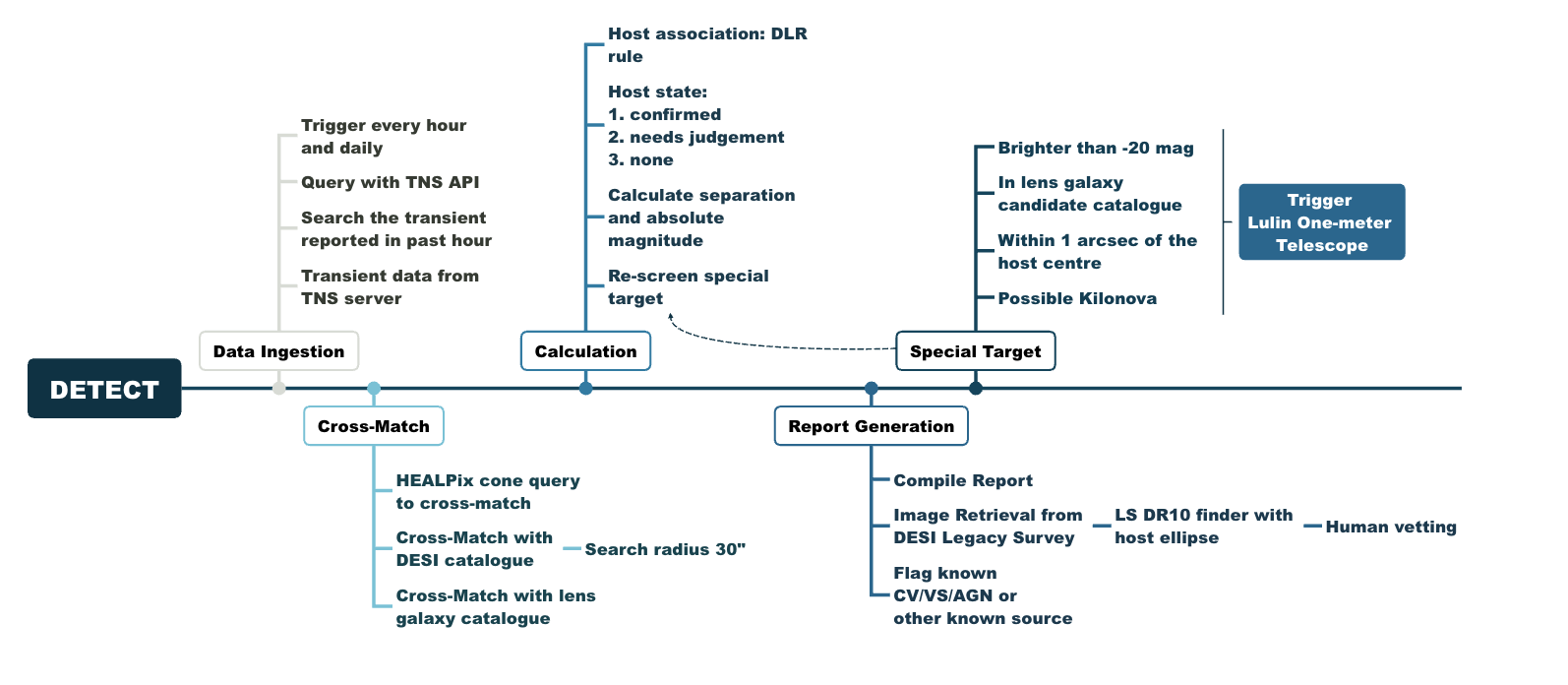}
    \caption{
    \textbf{Overview of the DETECT pipeline.}
    Every hour the pipeline ingests the TNS report of the previous hour (and once a day the daily file plus every object under follow-up), cross-matches each position against the $\sim$22 million DESI EDR+DR1 spectra (30\arcsec) and a strong-lens compilation (5\arcsec) through a HEALPix index, associates a host galaxy with the directional-light-radius rule of Section~\ref{sec:host_association}, and screens the event: known Galactic and AGN classifications are demoted, a peak absolute magnitude is derived from the light curve, and tags such as \emph{Luminous}, \emph{Too-bright}, \emph{Nuclear} and \emph{Lens} set a priority score. A Legacy Surveys DR10 finder with the host ellipse is produced and all results are written to the Kinder database. A reviewer sees each event with one of three host states (confirmed / needs judgement / none) and decides with one key whether it enters follow-up, that decision is kept on later runs.
    }
    \label{fig:detect_pipeline}
\end{figure*}

We developed \textbf{DETECT} as an automated pipeline to cross-match newly reported transients from the TNS with galaxies that have DESI spectroscopic redshifts. The pipeline is designed to provide candidate host associations and basic distance-related contextual information for rapid transient screening. A schematic overview of the workflow is shown in Figure~\ref{fig:detect_pipeline}.

\subsection{DESI Spectroscopic Catalogs}
\label{sec:desi_catalogs}
DESI is a large spectroscopic survey that measures redshifts for millions of galaxies and quasars \citep{2022AJ....164..207D}. We use the Early Data Release \citep[EDR,][]{2024AJ....168...58D} and Data Release 1 \citep[DR1,][]{desicollaboration2025datarelease1dark}. DR1 re-reduces the survey-validation and commissioning tiles that make up the EDR, so the catalogue consists of the $\sim$22 million primary DR1 spectra (\texttt{ZCAT\_PRIMARY}) plus the $\sim$22,000 EDR targets absent from DR1, joined to the Legacy Surveys DR10 Tractor photometry and shapes and to the DR1 stellar-mass and AGN value-added catalogues. All spectra are kept as candidates, but only reliable redshifts can be assigned as hosts: a spectrum carrying a Redrock \texttt{ZWARN} failure flag (NODATA, POORDATA, BAD\_TARGET, \ldots) is never assigned. \texttt{SPECTYPE=STAR} spectra are excluded from host association and used only to flag transients coincident with a Galactic starm, \texttt{QSO} spectra are accepted as hosts only for nuclear events (Section~\ref{sec:host_association}).

\subsection{Photometric Data}
For transients with a confirmed spectroscopic host, light curves are collected by the Kinder marshal from multiple sources. These include direct queries to ATLAS Force photometry \citep{Tonry_2018,2021TNSAN...7....1S,2020PASP..132h5002S}, Pan-STARRS \citep{2016arXiv161205560C, 2025MNRAS.542..541F}, Gaia \citep{2018A&A...616A..14G}, and ZTF \citep{2019PASP..131a8002B,2019PASP..131g8001G}, as well as photometry reported via TNS webpages from surveys such as WFST \citep{2023SCPMA..6609512W}, SDSS \citep{2000AJ....120.1579Y}, GOTO \citep{2022MNRAS.511.2405S}, BlackGEM \citep{2024PASP..136k5003G}, CRTS \citep{2009ApJ...696..870D}, and JWST/JADES \citep{2023arXiv230602465E}.

\subsection{Data Ingestion}
The pipeline is initiated automatically every hour. The primary data source utilized is the TNS, which serves as the official IAU mechanism for reporting new astronomical transients. Upon triggering, the system interfaces with the TNS Application Programming Interface (API) to execute a structured query, identifying and retrieving all transient candidates reported within the preceding window.

This process successfully ingests the raw transient data directly from the server, including essential metadata such as the target name, celestial coordinates (Right Ascension and Declination), discovery date, and discovery magnitude. Where available, existing spectroscopic classifications and redshifts reported by the community are also retained. This standardized dataset serves as the foundational input for the subsequent cross-matching stages.

\subsection{Cross Matching and Host Association}
\label{sec:crossmatch}

\subsubsection{Spatial cross-match}
Transient positions are matched against the DESI catalogue through a HEALPix index \citep[NESTED, $N_{\rm side}=8192$, $\sim$26\arcsec\ pixels,][]{2005ApJ...622..759G, 2019JOSS....4.1298Z}: the pixels covering a 30\arcsec\ cone are selected with an inclusive \texttt{query\_disc}, and the exact angular separation is applied afterwards, so the result is identical to a brute-force cone search at $\sim$10~ms per object. The same index serves the strong-lens compilation (Section~\ref{sec:lens_catalogs}) with a 5\arcsec\ radius, the scale of an Einstein radius. The 30\arcsec\ radius follows from the calibration sample of Section~\ref{sec:rule_calibration}: about 90\% of the true hosts of classified supernovae lie within $\sim$8\arcsec\ and nearly all within a projected $\sim$20~kpc.

\subsubsection{Host association}
\label{sec:host_association}
Each DESI spectrum within the cone is associated with a Legacy Surveys DR10 Tractor model \citep{2016ascl.soft04008L, 2019AJ....157..168D}, taking the model's half-light radius $R_{50}$, ellipticity components $(e_1, e_2)$ and profile type (EXP, DEV, REX or SER), when the live DR10 query fails we fall back to the shape stored in the catalogue. Following \citet{2006ApJ...648..868S} and \citet{2016AJ....152..154G} we define the directional light radius, the radius of the $R_{50}$ ellipse along the direction from the galaxy centre to the transient,
\begin{equation}
{\rm DLR} = \frac{ab}{\sqrt{(b\cos\phi)^2 + (a\sin\phi)^2}},
\end{equation}
where $a$ and $b$ are the semi-major and semi-minor axes and $\phi$ the angle between the major axis and that direction, and the normalised separation $d_{\rm DLR}=\theta/{\rm DLR}$ with $\theta$ the angular distance from the model centre. A floor of 0.5\arcsec\ on $R_{50}$ keeps unresolved sources from producing unphysically large $d_{\rm DLR}$. Spectra that fall on the same Tractor model, or whose redshifts agree to $\Delta z \le 0.001$ and whose centre lies inside the other's $R_{25}$ ellipse (shredded sources), are merged into one galaxy. Note that $R_{50}$ is a model half-light radius, whereas \citet{2006ApJ...648..868S} and \citet{2016AJ....152..154G} used isophotal or second-moment radii, the numerical threshold therefore cannot be transferred from those works and is calibrated here.

A galaxy is a host candidate (``member'') if $d_{\rm DLR} \le D_{\max}$, with $D_{\max}=4$ for de~Vaucouleurs, S\'ersic and composite models and $D_{\max}=8$ for exponential and round-exponential models, whose $R_{50}$ underestimates the extent of the disc (Section~\ref{sec:rule_calibration}), the host is the member with the smallest $d_{\rm DLR}$. Three situations are flagged for a reviewer instead of being decided automatically: (i) a second member within a factor 1.5 in $d_{\rm DLR}$ of the best (ambiguous), (ii) two spectra on one model with discrepant redshifts, typically a foreground--background blend or a DESI spectrum of the transient itself, (iii) a candidate with $D_{\max} < d_{\rm DLR} \le 2D_{\max}$ and no member (tentative). A model whose only spectrum is a quasar is accepted as host only for nuclear events ($d_{\rm DLR}\le1$), and a host that would imply $M\lesssim-23$ is flagged as a probable line-of-sight superposition. The thresholds and these special cases were derived from the visually vetted 2020--2024 sample of Section~\ref{sec:analysis}, which therefore serves both as the calibration set of the rule and as the sample analysed in Section~\ref{sec:results}.

\begin{figure}[htbp]
    \centering
    \includegraphics[width=1\linewidth]{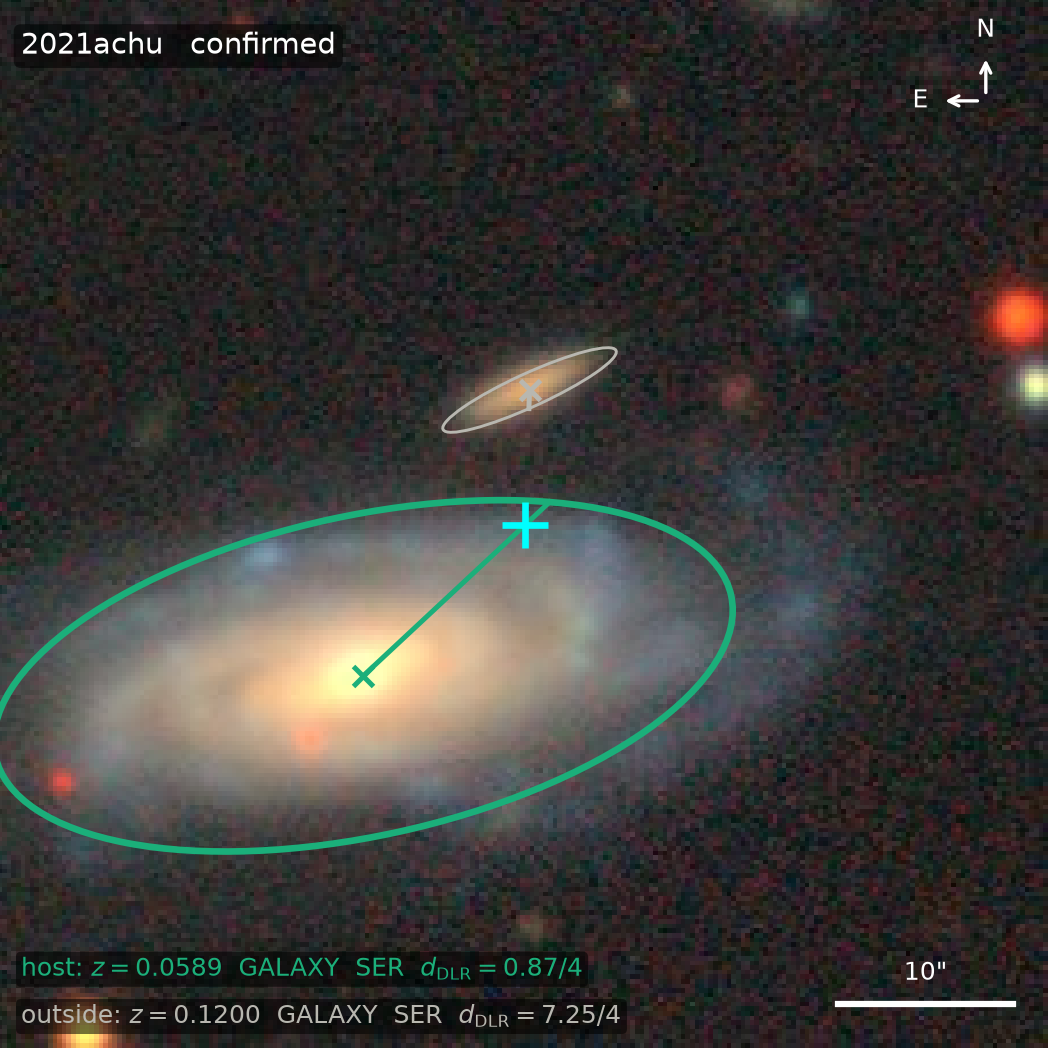}
    \caption{Legacy Surveys DR10 $grz$ cutout (north up, east left, the bar is 10\arcsec) with the transient as the cyan cross. Two DESI spectra lie within the search cone. The nearer one ($\sim$8\arcsec, grey) is a small galaxy at $z=0.120$ with $R_{50}\approx2\arcsec$: the transient is more than seven of its radii away ($d_{\rm DLR}\approx7$, beyond twice the threshold of 4), so it is not even a tentative host. The farther one ($\sim$13\arcsec, green) is the large spiral at $z=0.0589$ with $R_{50}\approx14\arcsec$: the transient sits inside its $R_{50}$ ellipse ($d_{\rm DLR}\approx0.9$). The green segment is the galaxy's DLR, its radius along the direction to the transient. A nearest-neighbour match would assign the $z=0.120$ galaxy and a distance modulus $\sim$1.6~mag too large, the rule assigns the spiral. Labels give the redshift, spectral and profile type, and $d_{\rm DLR}$ with its threshold.
    }
    \label{fig:chk_host}
\end{figure}

For each event a Legacy Surveys DR10 $grz$ cutout is produced with the host ellipse, its directional light radius toward the transient and any merged shreds, it is the first thing the reviewer sees. Figure~\ref{fig:chk_host} shows a typical case: the geometrically nearest spectrum is a small background galaxy, and the rule assigns the larger foreground spiral in whose light the transient lies. When the visually identified host has no DESI spectrum the event receives no redshift and is not part of the DESI-matched sample, matching against further spectroscopic catalogues is left for future work.

\subsubsection{Lens catalogues}
\label{sec:lens_catalogs}
Positions are also matched, within 5\arcsec, against a compilation of confirmed lenses and high-probability candidates: the Subaru HSC-SSP SuGOHI series \citep{2018PASJ...70S..29S, 2018ApJ...867..107W, 2019A&A...630A..71S, 2020A&A...636A..87C, 2020MNRAS.495.1291J, 2020A&A...642A.148S, 2021MNRAS.502.1487J, 2022PASJ...74.1209W, 2024MNRAS.527.6253C, 2024MNRAS.535.1625J}, HOLISMOKES \citep{2021A&A...653L...6C, 2022A&A...662A...4S, 2025A&A...693A.291S, 2025A&A...699A.350S}, and the machine-learning catalogues of \citet{2025arXiv250916033H} and \citet{2025arXiv251204275K}. The compilation is a secondary flag, not a search strategy: a coincidence raises the priority of an event, but lensed supernovae are expected to be found first through their luminosity (Section~\ref{sec:screening}).

\subsection{Absolute magnitude}
\label{sec:absmag}
The redshift adopted for an event is that of the host assigned by the rule of Section~\ref{sec:host_association}, or the classification redshift reported to TNS when no host is assigned. The apparent magnitude is the brightest real detection in the light curve stored for the object (upper limits and placeholder values excluded), falling back to the TNS discovery magnitude when no photometry has been collected; both versions are kept. The absolute magnitude is
\begin{equation}
M = m - \mu(z) - 2.5\log_{10}(1+z) - A_\lambda ,
\end{equation}
with the distance modulus $\mu$ for the flat $\Lambda$CDM cosmology of Section~\ref{sec:intro}, the bandwidth term of the $K$-correction \citep{2002astro.ph.10394H} without an SED-dependent term, and the Milky Way extinction $A_\lambda$ from the \citet{1998ApJ...500..525S} map through \texttt{dustmaps} \citep{2018JOSS....3..695M} with the \citet{2011ApJ...737..103S} coefficients for the reported filter. The same routine is used everywhere on the web interface, so a value quoted there is the value the pipeline computed. Objects under follow-up are re-screened once a day, so $M$ tracks the light curve as photometry accumulates.

\subsection{Screening, prioritization and review}
\label{sec:screening}
Every ingested event, matched or not, is screened once its host state is known. Known Galactic classifications on TNS (cataclysmic variables, variable stars, novae, YSOs) and known AGN (a TNS AGN/QSO class, or a nuclear event in a galaxy flagged by the DESI AGN catalogue or with a QSO spectrum) are demoted rather than removed, as is any object that already has a spectroscopic class, a DESI stellar spectrum within $\sim$1.5\arcsec\ is noted. Interest is expressed as tags that add to a priority score: \emph{Luminous} ($M\le-20$) and \emph{SLSN?} ($M\le-21$), \emph{Too-bright}, an event at $z\ge0.15$ more than one magnitude brighter than a normal SN~Ia at peak ($M\lesssim-20.3$), \emph{Nuclear} (within $\sim$1\arcsec\ of the host centre) and \emph{TDE?} (nuclear in a galaxy that is not an AGN), \emph{Passive-host} (${\rm sSFR}<10^{-11}\,{\rm yr^{-1}}$ or a de~Vaucouleurs profile), \emph{Lens} (a catalogued lens within 5\arcsec), \emph{glSN?} (too-bright in a passive host or on a lens) and \emph{Kilonova?} ($M>-17$ and a fade of $\ge0.5$~mag~day$^{-1}$ after the brightest detection, Section~\ref{sec:special}). The thresholds are deliberately permissive, the score orders the queue, it does not decide.

All results, the candidate rows, the host state (\emph{confirmed}, \emph{needs judgement}, or \emph{none}), the score and tags, the absolute magnitude and the finder, are written to the Kinder database and shown on the Kinder web interface.\footnote{\url{https://kinder.astro.ncu.edu.tw/detect}} There the reviewer sees each event with the pipeline's one-line verdict, the DR10 finder, the light curve and the candidate table, and decides with a single key whether the event enters follow-up, is closed, or has no host, a different candidate can be chosen as host in the same table. A reviewer's host choice is stored with the candidate and overrides the rule on every later run, and every event under follow-up is re-screened daily. The Kinder web application runs the same pipeline code after its own TNS import, so the two systems never disagree on a host or a magnitude.

\subsection{Special Target Identification and Follow-up Triggering}
\label{sec:special}
Beyond standard reporting, the pipeline includes a dedicated module to instantly identify physically rare or scientifically high-value targets based on specific photometric and environmental criteria. A target is flagged as ``Special'' if it meets any of the following conditions: (1) it exhibits an exceptionally high intrinsic luminosity (absolute magnitude $M < -20$ mag), suggesting a potential SLSN candidate, (2) it is spatially coincident with a known lens galaxy candidate from the catalogs, or (3) it possesses a small angular separation combined with a low (i.e., highly negative) absolute magnitude, indicative of a nuclear transient or a lensed event. Upon meeting these criteria, the system automatically triggers an alert for rapid follow-up observations with the Lulin One-meter Telescope (LOT), enabling the immediate acquisition of multi-band photometry or spectroscopy to confirm the nature of these transient events.

Specifically, the pipeline prioritizes the following classes of special targets for follow-up:
\begin{itemize}
    \item SLSNe: SLSNe often occur in faint dwarf galaxies and exhibit high intrinsic luminosities. Through DETECT monitoring, if a transient shows an anomalously bright absolute magnitude, while the host remains relatively faint, we flag it as an SLSN candidate. In practice, we adopt a relatively permissive threshold of $M<-20$ mag to trigger follow-up as early as possible, prioritizing completeness over purity at the alert stage. This immediately triggers follow-up optical photometry or spectroscopy using the Lulin Observatory.
    
    \item Lensed SNe: DETECT has no lens-specific selection. An event that is too luminous for its host redshift, more than one magnitude brighter than a normal SN~Ia at peak, $M\lesssim-20.3$ at $z\ge0.15$, is flagged whether or not a lens candidate lies nearby, the reviewer then examines the Legacy Surveys image for a lens-like configuration (a massive early-type galaxy, an arc, multiple images) and requests high-resolution imaging. A coincidence with a catalogued lens only raises the priority. SN~2025wny (Section~\ref{sec:wny}) was found this way, before lens catalogues were part of the pipeline, its lens galaxy is in the compilation now.

    \item Kilonovae: at the faint end, an unclassified event with $M\gtrsim-17$ whose light curve fades by more than $\sim$0.5~mag~day$^{-1}$ after its brightest detection, a least-squares slope over the following ten days, required to exceed the photometric errors, is tagged \emph{Kilonova?}, and its $g$, $r$ and $i$ photometry is compared with the \textsc{possis} kilonova grid of \citet{2020NatCo..11.4129C} on the marshal page. Because such events fade and redden within days, the tag triggers immediate multi-band follow-up. Applied to AT~2017gfo as if it were unclassified, the test tags it on the first day after peak.
    
    \item Host confusion: the three review states of Section~\ref{sec:host_association} (ambiguous, discrepant redshifts on one model, tentative) and the \emph{Unphysical-M} flag catch the cases where the nearest spectroscopic galaxy is not the host, SN~2023rgo (Section~\ref{sec:host_verification}) is the converse case, a background event behind a bright foreground galaxy, which only a spectrum of the transient itself can resolve.
\end{itemize}
The actual results of these real-time triggers, including detailed case studies of successfully identified candidates and subsequent vetting processes, are presented in Section~\ref{sec:results}.

% =============================================================
% =============================================================
% =============================================================

\section{Retrospective Analysis}
\label{sec:analysis}

To assess the performance and utility of DETECT, we carried out a retrospective analysis using archival data. In this section, we describe the construction of the historical sample, the data-preparation procedures used for large-scale processing, and the resulting host-galaxy recovery statistics.

\subsection{Historical Sample Selection (2020--2024)}
\label{sec:historical_sample}
We utilized the pipeline's Data Ingestion module to retrieve a complete dataset of transient events reported to the TNS between January 1st, 2020, and December 31st, 2024. This query yielded a total of $\sim1.2\times10^5$ entries. 

To ensure the purity of the sample for absolute magnitude analysis and to mimic the operational logic of the DETECT pipeline, we applied automated filtering criteria to remove non-SN contaminants. We excluded known Galactic cataclysmic variables (CVs) and variable stars (VS) by cross-matching with the VSX \citep{2006SASS...25...47W} and the CV catalogues of \citet{2001PASP..113..764D} and \citet{2003A&A...404..301R} via \texttt{Vizier} \citep{2000A&AS..143...23O} and \texttt{Astroquery} \citep{2019AJ....157...98G}. Additionally, known active galactic nuclei (AGN) and quasars were excluded to minimize contamination, thereby creating a cleaner dataset for the subsequent validation of host galaxy associations.

\subsection{Sky coverage}
\label{sec:sky_coverage}
\begin{figure*}[htbp]
    \centering
    \includegraphics[width=1\linewidth]{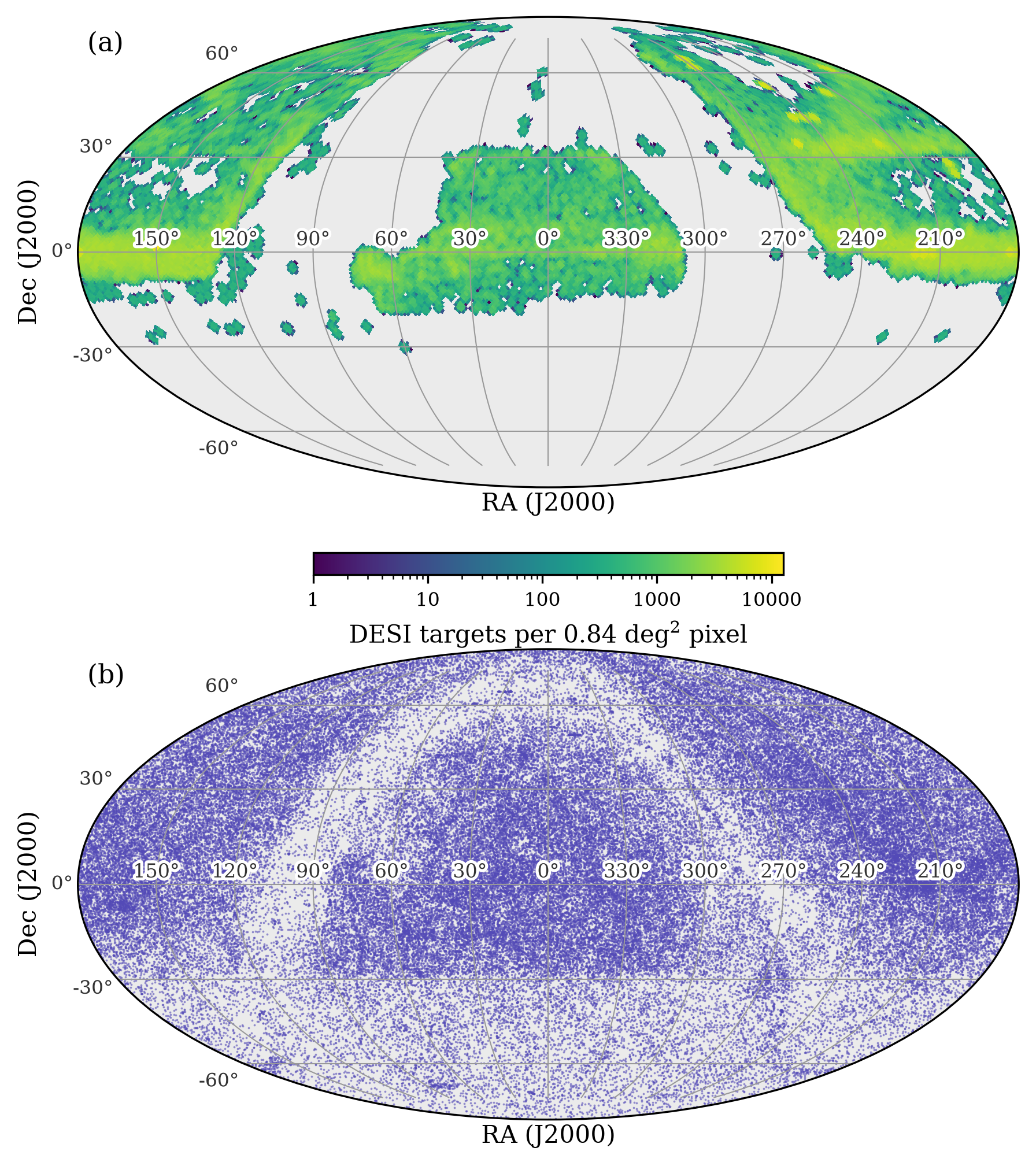}
    \caption{
        \textbf{Sky coverage of the two samples.}
        \textit{Top:} DESI EDR+DR1 targets on a HEALPix grid ($N_{\rm side}=64$, $\sim$0.8~deg$^2$ pixels), $\sim$22 million spectra over $\sim$15,000~deg$^2$.
        \textit{Bottom:} the $\sim1.2\times10^5$ TNS transients discovered in 2020--2024 on the same grid, about 59\% lie in pixels with DESI spectra. Mollweide projection in equatorial coordinates, logarithmic colour scale, grey where empty.
    }
    \label{fig:DESI_TNS_dis}
\end{figure*}

The retrospective run predates the database of Section~\ref{sec:crossmatch} and used a file-based RA/Dec grid, the two give identical matches. Figure~\ref{fig:DESI_TNS_dis} shows both samples on a HEALPix grid ($N_{\rm side}=64$, $\sim$0.8~deg$^2$ pixels): the DESI EDR+DR1 spectra cover $\sim$15,000~deg$^2$, and about 59\% of the TNS transients discovered in 2020--2024 fall in pixels with DESI spectra, which sets the ceiling of the host-recovery fraction.

\subsection{Host Verification and Redshift Reliability}
\label{sec:host_verification}
For the retrospective sample we used a generous 30\arcsec\ search radius, at the median redshift of classified supernovae, $z\approx0.05$, a 30~kpc stellar disc subtends $\sim$30\arcsec, so that events in the outskirts of large nearby galaxies were not lost. This yielded $\sim$39,000 transients with at least one candidate. Each was inspected on a Legacy Surveys cutout annotated with the five nearest spectroscopic galaxies, judging whether the transient lies on the galaxy's light, $\sim$17,600 events were confirmed, of which $\sim$16,800 remain after the removal of duplicates and mis-identified sources described below. This visually vetted sample is the training set of the host rule (Section~\ref{sec:rule_calibration}) and the parent sample of Section~\ref{sec:results}.

To validate the accuracy of these associations, we compared the DESI-derived redshifts with the spectroscopic redshifts reported on TNS for the subset of identified SNe. We found discrepancies in only 37 cases. A closer examination revealed that the majority of these mismatches arose from host confusion, where a background or foreground galaxy with a small angular separation from the true host was incorrectly targeted.

A notable example of such chance alignment resolved by our process is SN~2023rgo. This transient is located approximately $6.7^{\prime\prime}$ from the center of the bright foreground galaxy UGC~03416 ($z=0.0133$). A purely geometric cross-match would likely associate the transient with UGC~03416. However, follow-up spectroscopy revealed narrow galaxy emission lines at a much higher redshift of $z=0.226$, indicating that SN~2023rgo is actually a background source unrelated to the foreground galaxy \citep{2023TNSAN.274....1L}. Correcting for the true redshift shifts the derived absolute magnitude to $M_r \approx -20.8$, reclassifying the object from a seemingly faint transient in a local galaxy to an SLSN-I in the distant universe.

\subsection{Calibration of the host rule}
\label{sec:rule_calibration}
The rule of Section~\ref{sec:host_association} was calibrated on the $\sim$4,500 events of the vetted sample that are classified supernovae with a TNS redshift and have a DESI galaxy within a projected 30~kpc whose redshift agrees with it ($|\Delta z|\le0.01$), that galaxy is taken as the true host. The distribution of $d_{\rm DLR}$ of true hosts depends on the Tractor profile type: 95\% completeness is reached at $d_{\rm DLR}\approx4$ for DEV, SER and COMP models but only at $d_{\rm DLR}\approx8$ for EXP and REX models, whose half-light radius underestimates the disc. With these thresholds the rule recovers the true host for $\sim$93\% of the sample (about 86\% at $z<0.03$ and $\gtrsim95$\% at $z>0.06$, the loss at low redshift is transients in the outskirts of large nearby galaxies), assigns a wrong galaxy in $<1$\%, and assigns a host to $\sim$7\% of the events that have no redshift-consistent galaxy, $\sim$4\% of assignments are ambiguous. The result is insensitive to the floor imposed on $R_{50}$: raising it from 0 to 2\arcsec\ changes the completeness by under one percentage point, while the false-host rate rises from $\sim$7\% to $\sim$9\%. The previous criterion, containment within the $R_{25}$ ellipse, reaches $\sim$88\% completeness at a $\sim$5\% false-host rate.

Applied to the whole vetted sample, the rule agrees with the reviewer for $\sim$95\% of the events (the same galaxy, or another DESI spectrum of the same galaxy), it prefers a different galaxy for $\sim$70, mostly neighbours at nearly the same redshift, and declines to assign a host for $\sim$700, most of them tentative cases in which the reviewer accepted a galaxy just beyond the threshold. Figure~\ref{fig:chk_host} shows the most common way the rule and a nearest-neighbour match part company: the nearest spectrum belongs to a small galaxy at a higher redshift that the transient lies far outside of in units of its own radius, while a larger galaxy farther away on the sky contains it, the rule assigns the latter. The disagreements also exposed two failure modes now handled explicitly: spectra carrying the failed-fit redshift $z\simeq0.933$ (Section~\ref{sec:desi_catalogs}) and small background galaxies projected on the true host, which produce $M\lesssim-23$.

\subsection{Recovery Statistics}
\begin{figure}[htbp]
    \centering
    \includegraphics[width=1\linewidth]{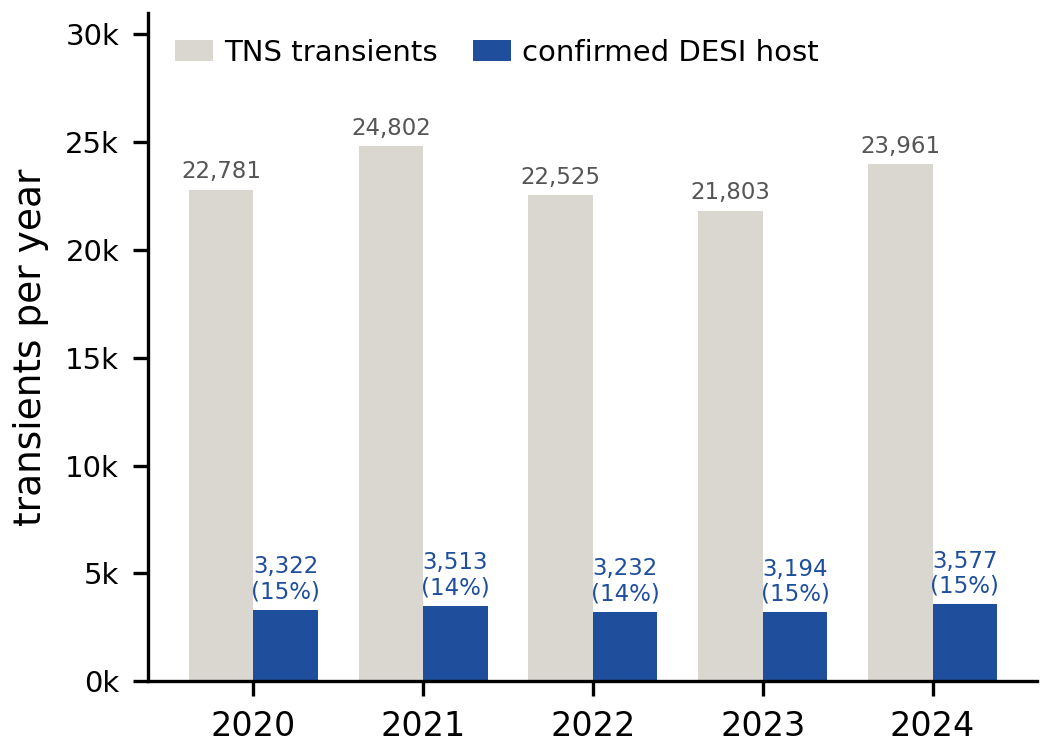}
    \caption{
        \textbf{Yearly recovery of the 2020--2024 TNS sample.}
        Grey: TNS transients per year after removing objects classified as Galactic (CV, variable star, nova) or AGN ($\sim$1.2$\times10^{5}$ in total). Blue: those with a visually confirmed DESI spectroscopic host ($\sim$16,800), with the fraction of the year's sample in parentheses. The fraction is $\sim$15\% in every year, about 59\% of the sample lies inside the DESI footprint and about a third has a DESI galaxy spectrum within 30\arcsec.
    }
    \label{fig:CM_results_with_TNS}
\end{figure}

The yearly distribution of the cross-matching results is presented in Figure~\ref{fig:CM_results_with_TNS}. In this comparison, the total height of each bar (gray) represents the TNS input sample, while the blue portion indicates the subset successfully associated with a DESI spectroscopic host after visual vetting. As described in Section~\ref{sec:historical_sample}, the parent sample was rigorously filtered to remove Galactic contaminants (CVs, VS) and explicit AGNs. However, it is important to acknowledge that this filtering relies on prior classifications, thus, a residual population of unclassified or quiescent AGNs may still exist within the dataset. Overall, the vetted cross-match fraction, the vetted sample divided by the filtered TNS sample, is about 15\% over the five years, roughly a quarter of the $\sim$59\% ceiling set by the DESI footprint (Section~\ref{sec:sky_coverage}).

% =============================================================
% =============================================================
% =============================================================

\section{Results}
\label{sec:results}

\subsection{Retrospective Results}
Using the DESI-matched ``Gold Sample'' constructed from the 2020--2024 archival data (Section~\ref{sec:analysis}), we now characterize the physical properties of
the recovered transients and demonstrate how host-redshift information enables the
identification of rare and extreme events. We first derive robust peak absolute
magnitudes through parameterized light-curve fitting
(Section~\ref{sec:abs_mag_analysis}) and examine the resulting luminosity
distributions across spectroscopic subtypes. We then combine these luminosities
with host-galaxy properties to isolate SLSN candidates via a
purely photometric high-contrast selection (Section~\ref{sec:host_transient}), and
finally apply the framework to screen for fast-evolving and gravitationally lensed
transients (Section~\ref{sec:rare_transient}).

\subsubsection{Absolute Magnitude Analysis}
\label{sec:abs_mag_analysis}
To mitigate the underestimation of peak magnitudes caused by observational cadence gaps and to derive more physically accurate peak magnitudes across all filters, we employed the \texttt{Haffet} \citep{2021A&A...655A..90Y} package to perform parameterized fitting on the multi-band light curves.

The data preprocessing pipeline proceeded as follows: First, adopting the ATLAS Force Photometry format as our baseline, we converted all relative magnitude measurements into flux units. This approach was chosen because flux space maintains better statistical linearity in low signal-to-noise ratio (SNR) regimes. Next, to enhance data purity, we applied a two-stage cleaning process to the data in each filter: first, we utilized a rolling median method to identify and reject statistical outliers, second, we removed detection points with an SNR $< 3$ to eliminate background noise interference. For repeated observations obtained within the same night and filter, we merged them into a single epoch using error-weighted averaging (stacking). This step not only reduced data noise but also smoothed out short-timescale photometric fluctuations. For the ATLAS-specific $c$ (cyan) and $o$ (orange) filters, we adhered to the standard processing and merging procedures of the code from \citet{Young_plot_atlas_fp}.

Upon obtaining the cleaned daily flux light curves, we implemented a data quantity cut. To ensure sufficient degrees of freedom for the model fitting, we retained only those targets possessing at least 5 independent observation points in a single filter. This selection criterion reduced the sample size to 10,450 entries, which were then processed using \texttt{Haffet}.

Finally, to ensure the robustness of the derived physical parameters, we performed a rigorous post-fitting Quality Control (QC). We excluded cases where the derived peak magnitude had an SNR $< 3$, which typically indicates a failure of the fit to converge effectively. Furthermore, to prevent model over-extrapolation in data-sparse temporal regions, we mandated that the fitted peak time must fall within the temporal coverage of the data (i.e., requiring observation points both before and after the peak). Following these stringent vetting procedures, we established a high-confidence ``Gold Sample,'' comprising 5,400 transients ($\sim$10,600 single-filter light curves): 3,569 unclassified transients and 1,831 spectroscopically confirmed SNe. This `Gold Sample' forms the basis for the detailed peak absolute-magnitude distribution analysis presented in subsequent sections, ensuring that derived physical properties are robust against photometric uncertainties.

\begin{figure}[htbp]
    \centering
    \includegraphics[width=\columnwidth]{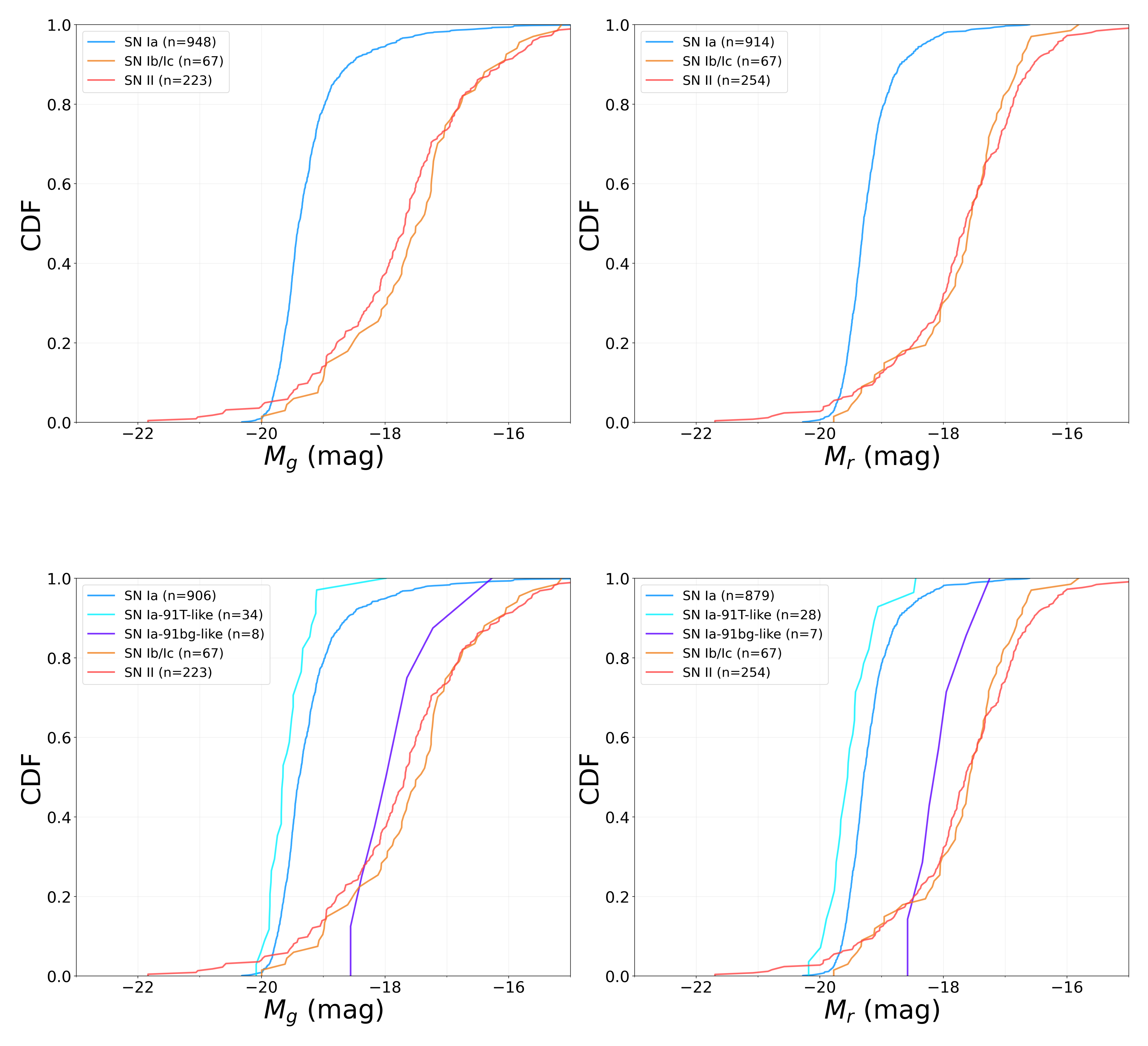}
    \caption{
        \textbf{Cumulative Distribution Functions (CDFs) of fitted peak absolute magnitudes for spectroscopically classified SNe in the `Gold Sample'.} 
        The left panel shows distributions in the $g$-band, and the right panel in the $r$-band. The colors indicate different subtypes, with sample sizes reported as ($g$-band/$r$-band) counts: normal SN~Ia (blue, n=906/879), SN~Ia-91T-like (cyan, n=34/28), SN~Ia-91bg-like (purple, n=8/7), SN~Ib/Ic (orange, n=67), and SN~II (red, n=223/254). The steep slope of the normal SN~Ia population signifies their low luminosity dispersion. Core-collapse SNe (Ib/Ic and II) display broader distributions extending to fainter magnitudes, while SN~Ia-91T-like events constitute the brightest subgroup.
    }
    \label{fig:SN_absolute_mag_cdf}
\end{figure}

To compare the luminosity characteristics across different SN classes within our spectroscopically confirmed `Gold Sample', we plotted the cumulative distribution functions (CDFs) of their fitted peak absolute magnitudes in the $g$- and $r$-bands (Figure~\ref{fig:SN_absolute_mag_cdf}). The sample is subdivided into five distinct spectroscopic subtypes based on TNS classifications: normal SN~Ia, overluminous SN~Ia-91T-like, subluminous SN~Ia-91bg-like, stripped-envelope core-collapse SN~Ib/Ic, and hydrogen-rich core-collapse SN~II. The CDF visualization clearly illustrates the physical differences between these populations. Normal SN~Ia exhibit the steepest rise in cumulative probability, reflecting their characteristic low luminosity dispersion as standardizable candles. In contrast, the core-collapse classes (SN~II and SN~Ib/Ic) show significantly broader distributions extending towards fainter magnitudes. Notably, the SN~Ia-91T-like subgroup consistently represents the brightest population in both filters, while SN~Ia-91bg-like events are clustered among the faintest.

\begin{figure}[htbp]
    \centering
    \includegraphics[width=\columnwidth]{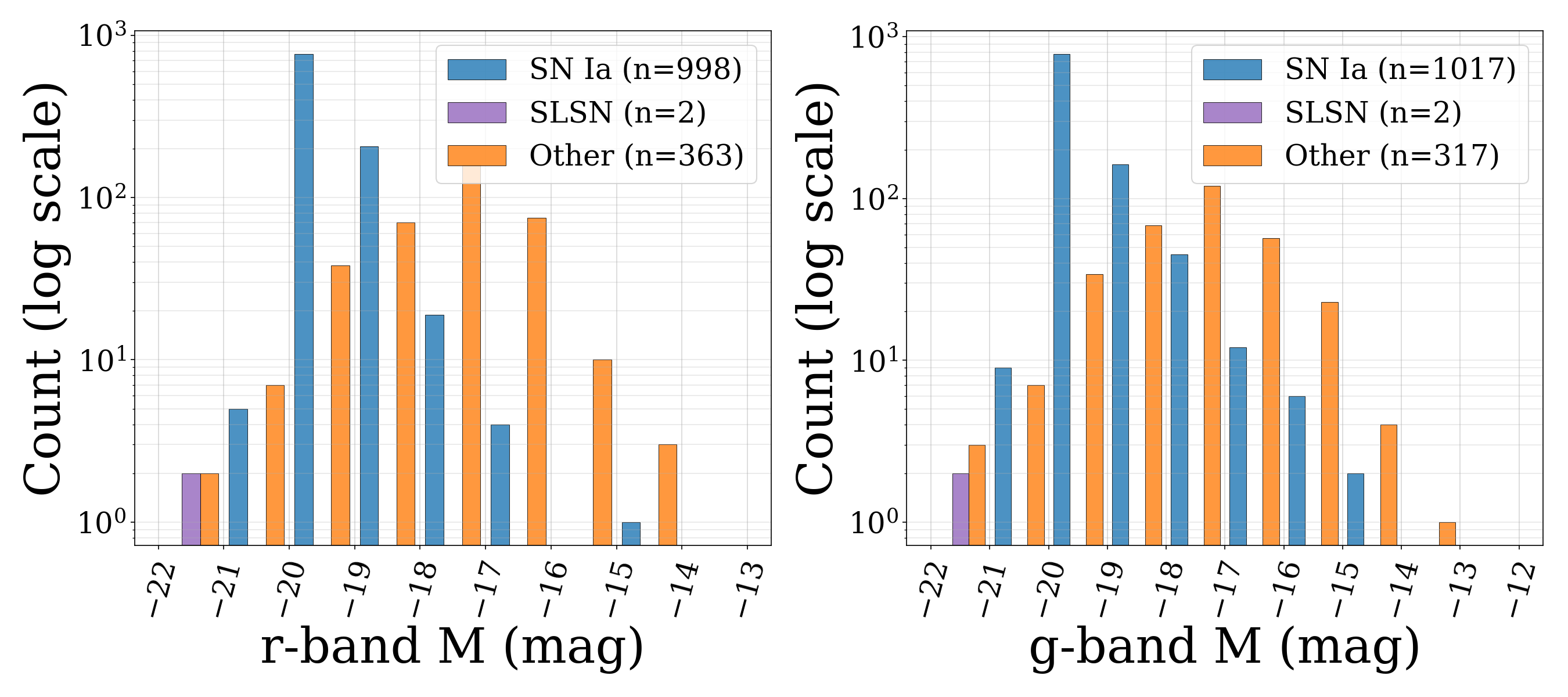}
    \caption{
        \textbf{Distribution of fitted peak absolute magnitudes for the Gold Sample.} 
        The histograms display the number of events per magnitude bin for the $r$-band (left panel) and $g$-band (right panel). 
        The sample is classified into three categories: Type Ia SNe (blue), SLSNe (purple), and other types (orange, primarily core-collapse SNe and unclassified transients). 
        Note that the count axis (y-axis) is on a logarithmic scale. 
        The distribution highlights the standard candle nature of SN~Ia, peaking narrowly around $M \approx -19$, whereas the ``Other'' population exhibits a broader distribution extending to fainter magnitudes. 
        SLSNe are distinctly located at the high-luminosity extreme ($M < -21$).
    }
    \label{fig:SN_absolute_mag_cdf_bar}
\end{figure}

\subsubsection{Host Galaxy and Transient}
\label{sec:host_transient}
For the unclassified transients we first ask whether the event could be nuclear activity rather than a supernova. For every host we take the WISE \citep{2010AJ....140.1868W} W1--W4 fluxes provided with the Legacy Surveys DR10 catalogue (unWISE forced photometry at the optical positions, \citealt{2019AJ....157..168D}). These fluxes are on the AB system, because the mid-infrared selection criteria are defined in Vega magnitudes, we convert them with the standard offsets $m_{\rm AB}-m_{\rm Vega}=2.699$, 3.339, 5.174 and 6.620 for W1--W4 \citep{2011ApJ...735..112J}, where W1, W2, W3 and W4 are the 3.4, 4.6, 12 and 22~$\mu$m bands.

\begin{figure}[htbp]
    \centering
    \includegraphics[width=1\linewidth]{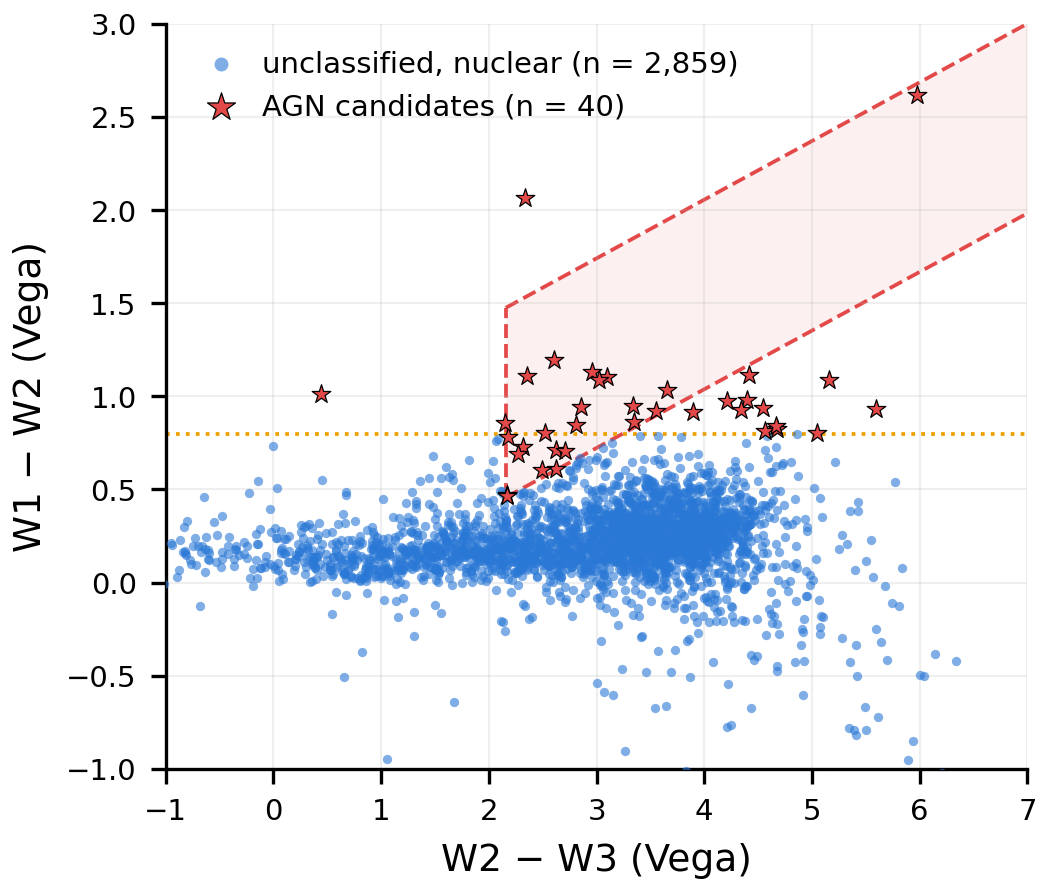}
    \caption{
    \textbf{WISE colour--colour diagram of the hosts of unclassified nuclear transients.}
    Vega colours, from the Legacy Surveys DR10 unWISE fluxes with the standard AB$-$Vega offsets, for the $\sim$2,900 unclassified TNS transients that lie within 1\arcsec\ of the DR10 model centre of their DESI host (blue). The shaded region is the AGN wedge of \citet{2012MNRAS.426.3271M} and the dotted line the \citet{2012ApJ...753...30S} criterion $W1-W2\geq0.8$, the transients whose host satisfies either criterion are flagged as AGN candidates (red stars, 40 of the 58 have a W3 detection and appear here) and are not treated as supernova candidates.
    }
    \label{fig:wise_color_distribution}
\end{figure}

A host is taken to be a mid-infrared AGN if it lies inside the \citet{2012MNRAS.426.3271M} wedge, $W2-W3>2.157$ and $0.315\,(W2-W3)-0.222<W1-W2<0.315\,(W2-W3)+0.796$, or satisfies the \citet{2012ApJ...753...30S} criterion $W1-W2\geq0.8$. The host colour alone does not identify the transient: six of the $\sim$390 spectroscopically classified supernovae found within 1\arcsec\ of their host centre, among them two SLSNe, have hosts that meet these criteria. We therefore combine the colour with the position of the transient and consider as AGN candidates only the unclassified transients that lie within 1\arcsec\ of the DR10 model centre of a WISE-selected AGN host. Figure~\ref{fig:wise_color_distribution} shows the $\sim$2,900 unclassified nuclear transients with WISE colours, 58 of them (about 1.4\%) meet the AGN criteria and are removed from the supernova candidate pools of the following sections. None of them enters the high-contrast selection below. Only two of the 40 hosts are classified as QSO by the DESI pipeline, the others having ordinary galaxy spectra, so the mid-infrared colour adds information that the optical spectrum does not.

Following the mid-infrared vetting, we examined the optical properties by comparing the host galaxy magnitudes with the transient luminosities. The $r$-band magnitude of each confirmed host is the Legacy Surveys DR10 Tractor model magnitude joined to its DESI spectrum (Section~\ref{sec:desi_catalogs}), corrected for Galactic extinction, its absolute magnitude $M_{r,\rm host}=m_r-\mu(z)-2.5\log_{10}(1+z)$ uses the same distance modulus and bandwidth term as the transients (Section~\ref{sec:absmag}) and is compared with the fitted peak absolute magnitude $M_{r,\rm AT}$.

\begin{figure}[htbp]
    \centering
        \includegraphics[width=\columnwidth]{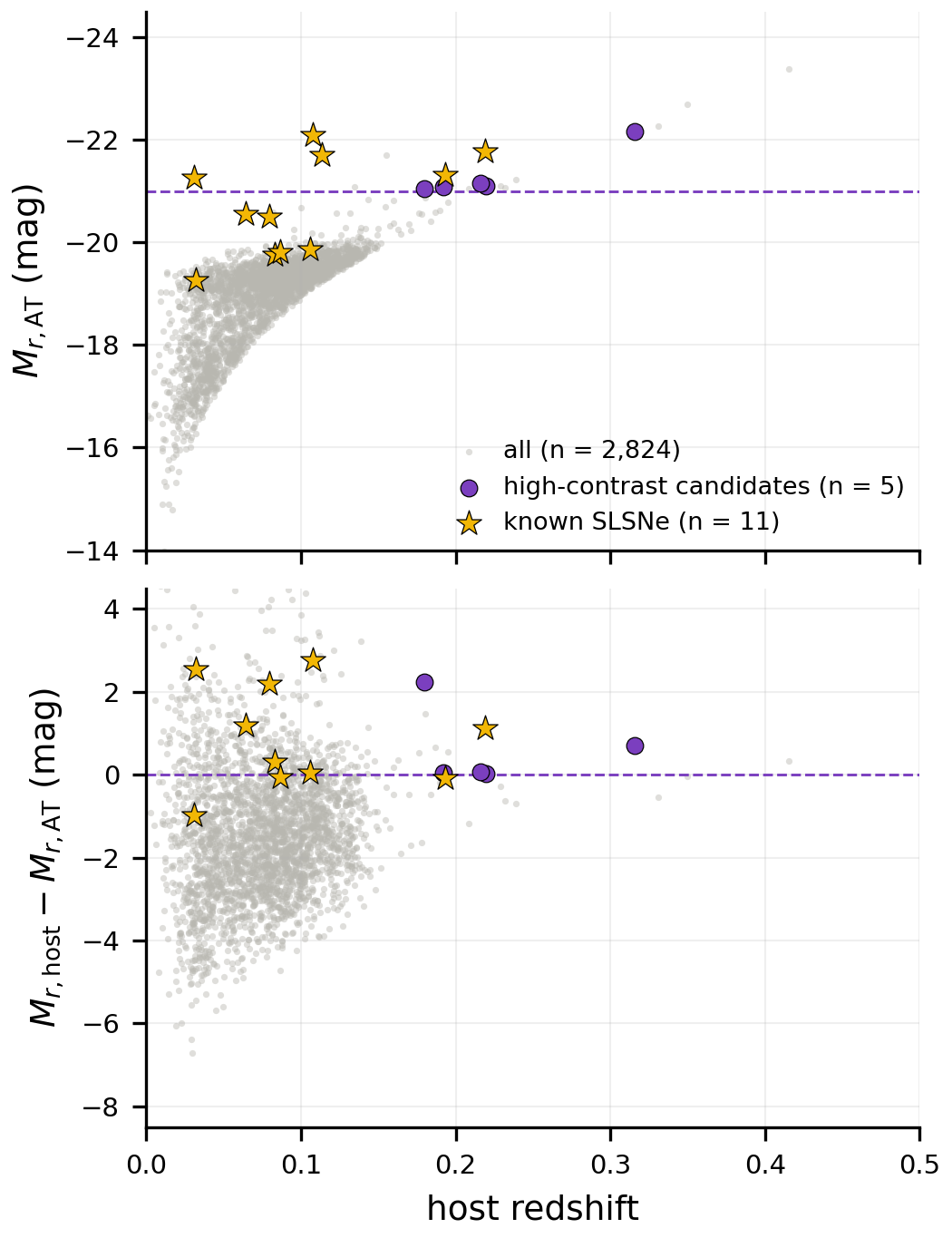}
    \caption{
        \textbf{Transient and host absolute magnitudes in the $r$ band.}
        \textit{Top:} fitted peak absolute magnitude of the transient, $M_{r,\rm AT}$, against the host redshift for the $\sim$2,800 Gold Sample transients with an $r$-band fit and a host magnitude.
        \textit{Bottom:} the difference $M_{r,\rm host}-M_{r,\rm AT}$, positive values mean the transient outshines its host. Dashed lines mark the selection $M_{r,\rm AT}<-21$ and $M_{r,\rm host}-M_{r,\rm AT}>0$.
        Gold stars: known SLSNe (the two spectroscopically classified ones in the sample and nine from the literature). Purple circles: the five unclassified high-contrast candidates of Table~\ref{tab:slsn_candidates}. Three events that pass the cuts but are excluded as WISE AGN candidates or as unphysically bright ($M<-23$) are left in grey; two of them, at $z>1$, are outside the plotted range.
    }
    \label{fig:hm_plot_r_filter}
\end{figure}

The distribution of these events is presented in Figure~\ref{fig:hm_plot_r_filter}. The upper panel displays the transient peak absolute magnitude ($M_{r,\rm AT}$) as a function of redshift, while the lower panel shows the magnitude difference ($M_{r,\rm host} - M_{r,\rm AT}$). In this definition, a positive difference ($M_h - M_t > 0$) indicates that the transient is intrinsically brighter than its host galaxy.

To isolate potential high-interest targets, specifically SLSNe, which are known to exhibit high intrinsic luminosities and often reside in faint dwarf galaxies \citep{2017ApJ...849L...4C}, we defined a high-contrast selection region. This selection applies two strict criteria. The transient must be exceptionally bright, with a peak absolute magnitude $M_{r,\rm AT} < -21$ mag from \citet{2012Sci...337..927G}. The transient must outshine its host galaxy, satisfying the condition $M_{r,\rm host} - M_{r,\rm AT} > 0$.

Figure~\ref{fig:hm_plot_r_filter} shows the $\sim$2,800 Gold Sample transients with an $r$-band fit and a host magnitude. Known SLSNe (gold stars: the two in the sample, SN~2021ek and SN~2024ixo, and nine from the literature) populate the high-luminosity, positive-contrast corner, as expected for events in faint hosts. Five unclassified transients meet both criteria and are neither WISE AGN candidates (Section~\ref{sec:host_transient}) nor unphysically bright (purple circles, Table~\ref{tab:slsn_candidates} in Appendix~\ref{sec:appendix_candidates}). Of the five, only AT~2020mmc outshines its host by more than two magnitudes, the other four sit within 1\arcsec\ of the centre of a host as luminous as themselves ($M_{r,\rm host}\approx-21$), a configuration in which nuclear activity is as plausible as a superluminous supernova. They are candidates for follow-up verification, to distinguish intrinsic superluminous events, gravitationally lensed supernovae and AGN flares, rather than a clean SLSN sample. The remaining population is shown in grey. This purely photometric selection separates the extreme tail from the bulk of normal supernovae without model predictions, but, with host magnitudes of this quality, it does not by itself separate SLSNe from nuclear transients in luminous hosts.

In addition to the global properties of the host galaxies, the location of a transient within its host provides critical constraints on its progenitor system. We calculated the projected physical offset ($d_{\rm proj}$) for each transient relative to the host galaxy center. 
The physical offset is derived from the angular separation $\theta$ (in arcseconds) obtained from our cross-matching process and the angular diameter distance $d_A(z)$ at the host's spectroscopic redshift.
Crucially, $d_A(z)$ is calculated assuming the same flat $\Lambda$CDM cosmology ($H_0 = 67.4\,\mathrm{km}\,\mathrm{s}^{-1}\,\mathrm{Mpc}^{-1}$, $\Omega_{\mathrm{m}} = 0.315$).

\begin{figure}[htbp]
    \centering
    \includegraphics[width=1\linewidth]{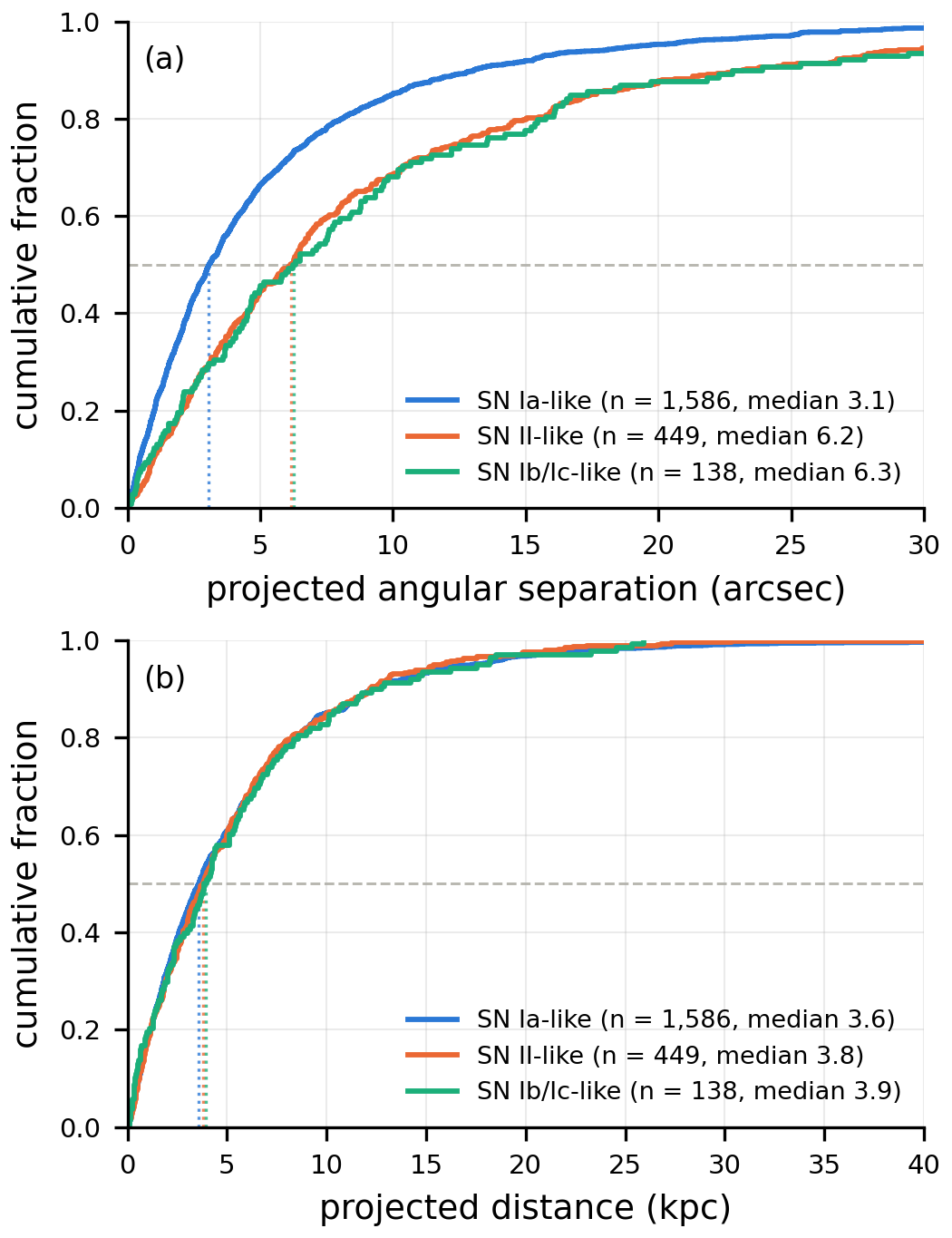} 
    \caption{
    \textbf{Cumulative distributions of the offset from the host centre for supernovae in the vetted sample.}
    \textit{(a)} Angular separation and \textit{(b)} projected distance at the host's DESI redshift, for SN~Ia-like (blue, n$\,\approx\,$1,600), SN~II-like (orange, n$\,\approx\,$450) and SN~Ib/Ic-like (green, n$\,\approx\,$140) events, dotted lines mark the medians. The angular offsets of the core-collapse families are twice those of SNe~Ia because their hosts are nearer, in physical units the three families coincide.
    } 
    \label{fig:cdf_offsets}
\end{figure}

Figure~\ref{fig:cdf_offsets} shows the cumulative distributions of the offset between each supernova and its host centre for the $\sim$2,200 TNS-classified supernovae in the vetted sample, in three families: SN~Ia-like (n$\,\approx\,$1,600), SN~II-like (n$\,\approx\,$450) and SN~Ib/Ic-like (n$\,\approx\,$140). In angular terms the families differ by a factor of two (Figure~\ref{fig:cdf_offsets}a): median separations of $\sim$3\arcsec\ for SNe~Ia against $\sim$6\arcsec\ for both core-collapse families, and 90th percentiles of $\sim$13\arcsec\ against $\sim$23\arcsec. This is a redshift effect. The core-collapse events are found in nearer galaxies (median $z\approx0.03$, against $\approx0.06$ for SNe~Ia), where the same physical offset subtends twice the angle, converted to projected distance at the host redshift (Figure~\ref{fig:cdf_offsets}b) the three distributions coincide, with medians of $\sim$3.6, $\sim$3.8 and $\sim$3.9~kpc and 90th percentiles of $\sim$12~kpc for all three. A fixed angular search radius therefore penalises nearby core-collapse hosts most, which is one motivation for the radius-normalised rule of Section~\ref{sec:host_association}.

We do not interpret the residual differences between the families physically. In the literature SNe~Ia extend to larger offsets than core-collapse events and SNe~Ib/c are the most centrally concentrated \citep[e.g.][]{2012ApJ...759..107K, 2014A&A...572A..38G}, the ATLAS100 volume-limited sample \citep{2026MNRAS.549g1028S} finds 90th-percentile offsets of $\sim$12~kpc for SNe~Ia and $\sim$8~kpc for stripped-envelope events. Our sample is limited to hosts targeted by DESI, bright, massive galaxies, and to a fixed angular radius, which removes the outskirt and intracluster events that separate the families, the agreement of our 90th percentiles with the ATLAS100 SN~Ia value shows that within this selection the offset scale is recovered.

\subsubsection{Identification of Rare Transients}
\label{sec:rare_transient}
In addition to identifying intrinsically bright candidates like SLSNe or lensed SNe, our pipeline is also designed to screen for rapidly evolving and fainter transients, such as kilonovae. To assess the coverage of our sample in the kilonova parameter space, we compared the derived absolute magnitudes of all unclassified transients (ATs) with theoretical kilonova model grids from \citet{2020NatCo..11.4129C}. These models were computed using the time-dependent Monte Carlo radiative transfer code \textsc{possis} \citep{2019MNRAS.489.5037B}, covering a wide range of ejecta masses, velocities, and viewing angles.

Furthermore, we included the observational data of the best-studied electromagnetic counterpart to a gravitational wave merger, AT~2017gfo, to serve as a fiducial baseline for identifying potential kilonova candidates \citep{2017PASA...34...69A, 2017Natur.551...64A, 2017ApJ...848L..19C, 2017ApJ...848L..17C, 2017Sci...358.1570D, 2017Sci...358.1565E, 2017Sci...358.1559K, 2017Natur.551...67P, 2017Natur.551...75S, 2017ApJ...848L..16S, 2017ApJ...848L..27T, 2017Natur.551...71T, 2017PASJ...69..101U, 2017ApJ...848L..24V}. By overlaying these theoretical and observational baselines, we can effectively distinguish potential KN-like events from the bulk of the SN population.

Through the systematic screening of unclassified transients against the kilonova phase space defined by the \textsc{possis} models and AT~2017gfo, our pipeline flagged two notable candidates: AT~2020bgp and AT~2020abut.

\begin{figure}[htbp]
    \centering
    \includegraphics[width=1\linewidth]{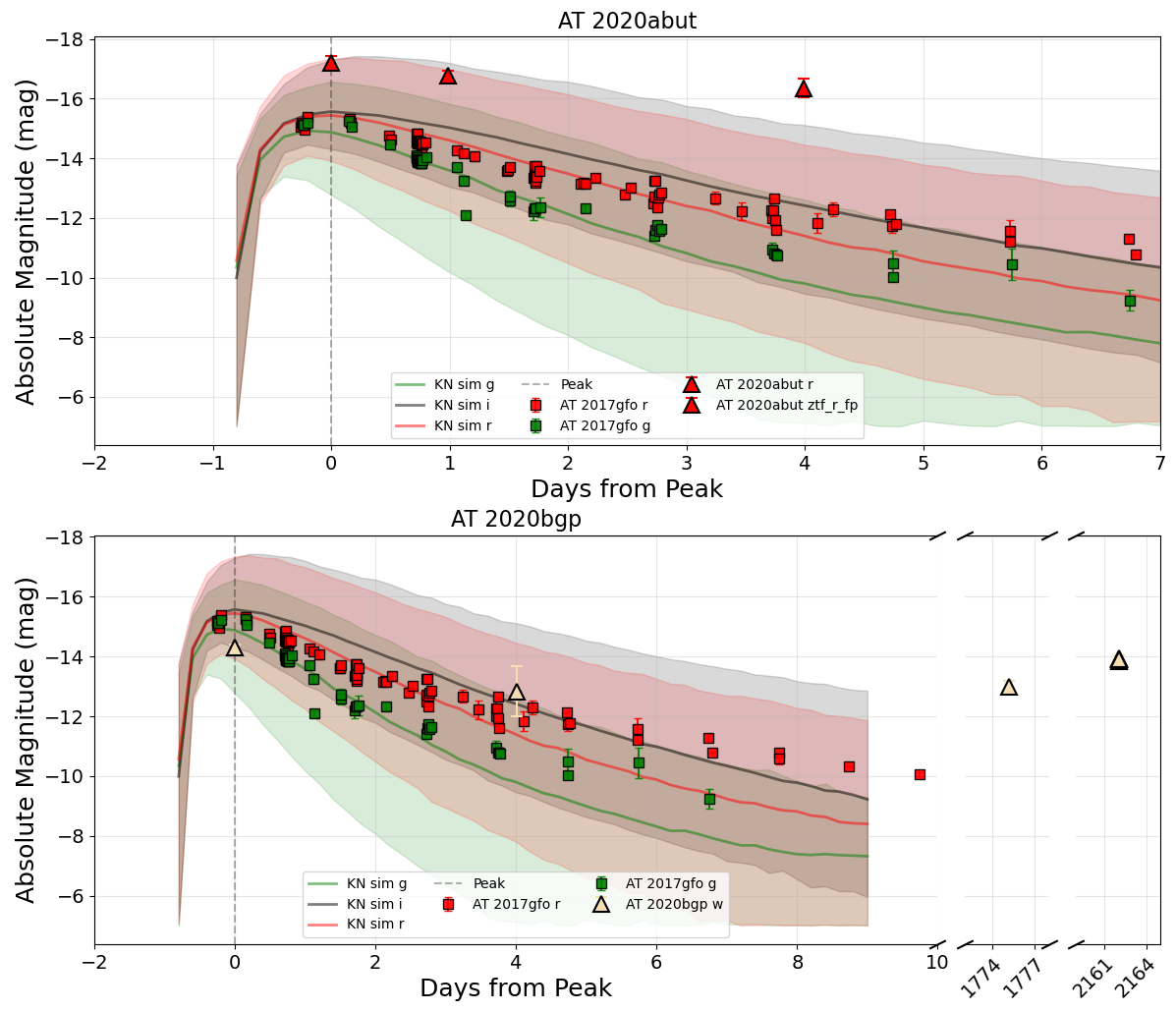}
    \caption{
        \textbf{Light curve comparison of the two fast-evolving transient candidates flagged by DETECT against kilonova model grids.}
        The shaded regions show the kilonova model grids from \citet{2020NatCo..11.4129C} in $g$ (green), $r$ (red), and $i$ (grey) bands, with the AT~2017gfo observational data overplotted as squares for reference.
        \textit{Top:} AT~2020abut (red triangles, $r$-band detections and ZTF forced-photometry points) declines more slowly than the kilonova model envelope at $t \gtrsim 1$\,day, inconsistent with a kilonova origin once the deeper ZTF forced photometry is included.
        \textit{Bottom:} AT~2020bgp ($w$-band, light triangles) tracks the kilonova model grid and the AT~2017gfo baseline closely over the first $\sim$9 days post-peak. However, archival Pan-STARRS photometry reveals two additional, fainter detections at $\approx1774$--$1777$ and $\approx2161$--$2164$ days after peak(right panels, separated by axis breaks), corresponding to recurrent outbursts roughly 4.8 and 5.9 years after the original event. This repeated, non-fading behavior rules out a kilonova interpretation for AT~2020bgp and instead points to a recurrent stellar eruption, such as a luminous blue variable (LBV) outburst or other supernova-impostor event.
    }
    \label{fig:kn_candidates}
\end{figure}

Figure~\ref{fig:kn_candidates} presents the absolute magnitude evolution of these two targets superimposed on the theoretical model grids. AT~2020bgp initially exhibits a rapid luminosity decline highly consistent with the kilonova evolutionary track over the first $\sim$9 days post-peak, with photometric data points falling well within the predicted bounds of the theoretical models and closely mirroring the decay rate of AT~2017gfo. This object was previously identified and reported as a potential kilonova candidate by \citet{2020TNSAN..31....1S}, and its recovery by DETECT based on the early-time photometry alone demonstrates the sensitivity of our automated search algorithm to fast-evolving, intrinsically faint transients.

However, archival Pan-STARRS photometry reveals two additional detections at $\approx1774$--$1777$ and $\approx2161$--$2164$ days after peak, corresponding to recurrent outbursts approximately 4.8 and 5.9 yr after the original event (Figure~\ref{fig:kn_candidates}, right panels). A genuine kilonova, powered by the radioactive decay of $r$-process ejecta from a compact-object merger, is by construction a single, non-repeating event, the presence of multiple, comparably faint outbursts from the same position is therefore incompatible with a kilonova origin. We instead interpret AT~2020bgp as a recurrent stellar eruption, most plausibly a luminous blue variable (LBV) outburst or related ``supernova impostor'' event \citep{2025MNRAS.542..541F}. Such eruptions, which arise from non-terminal mass-loss episodes in massive stars rather than explosive disruption, are known to repeat on timescales of years to decades and have been identified as a significant source of kilonova-candidate contamination in untargeted optical searches, accounting for roughly half of the false positives in the Pan-STARRS Search for Kilonovae sample. This case illustrates that, while DETECT's photometric criteria correctly flagged AT~2020bgp as a faint and rapidly evolving event worthy of follow-up, deep archival photometry spanning multiple years is essential to distinguish a genuine kilonova from a recurrent stellar impostor.

Similarly, AT~2020abut was initially flagged by our system due to its rapid photometric evolution based on the sparse public data. While intrinsically brighter than the standard model grid for a typical viewing angle, its apparent fast decline rate initially placed it in the high-interest region. However, subsequent verification using ZTF forced photometry \citep{2023arXiv230516279M} revealed additional detections that are inconsistent with a kilonova evolution. Consequently, we exclude AT~2020abut as a viable kilonova candidate. Together with the AT~2020bgp case above, this underscores the importance of deep, archival photometric validation (e.g., forced photometry, multi-epoch survey coverage) to refine candidates identified by automated pipelines and to distinguish genuine kilonovae from astrophysical impostors.

The complete list of these candidates is provided in Table~\ref{tab:kilonova_candidates} in Appendix~\ref{sec:appendix_kilonova}.

The luminosity-based selection above is DETECT's route to lensed supernovae, as a consistency check we also cross-matched the unclassified 2020--2024 transients against the lens catalogues of \citet{2025arXiv250916033H} and \citet{2025arXiv251204275K}. Five events lie within $\sim$2\arcsec\ of a catalogued lens (Table~\ref{tab:lensed_candidates} in Appendix~\ref{sec:appendix_lens}), with the DESI redshift of the lens they are worth re-examination, but none of them was flagged by luminosity, and we list them as positional coincidences rather than candidates.

\subsection{Real-Time Discoveries and Follow-up Validation}
A primary goal of DETECT is the early identification of rare or unusual transients for rapid follow-up. Since 2025 January, the monitoring pipeline has identified several such targets, enabling prompt validation and follow-up observations.

DETECT commenced its operational runs in early 2025, initially leveraging the DESI EDR as the primary spectroscopic reference. Following the public release of DESI DR1 in March 2025, the pipeline's internal database was promptly updated to incorporate these millions of additional spectra. This expansion significantly increased the volume of searchable host galaxies and improved the recovery rate of spectroscopic redshifts for newly reported transients.

The spatial cross-matching module was systematically augmented following the real-time discovery and subsequent confirmation of the gravitationally lensed supernova SN~2025wny in September 2025. Its lens galaxy was already a catalogued candidate, that the event was nevertheless found by luminosity alone, and would have been flagged twice had the catalogue been in place, motivated adding the lens compilation as a secondary flag. Consequently, we have since iteratively integrated multiple established lens catalogs into the DETECT workflow to maximize the detection potential for future lensed transients and other extreme events.

\subsubsection{SN~2025wny: Lensed SLSN}
\label{sec:wny}
\begin{figure}[htbp]
    \centering
    \includegraphics[width=0.9\linewidth]{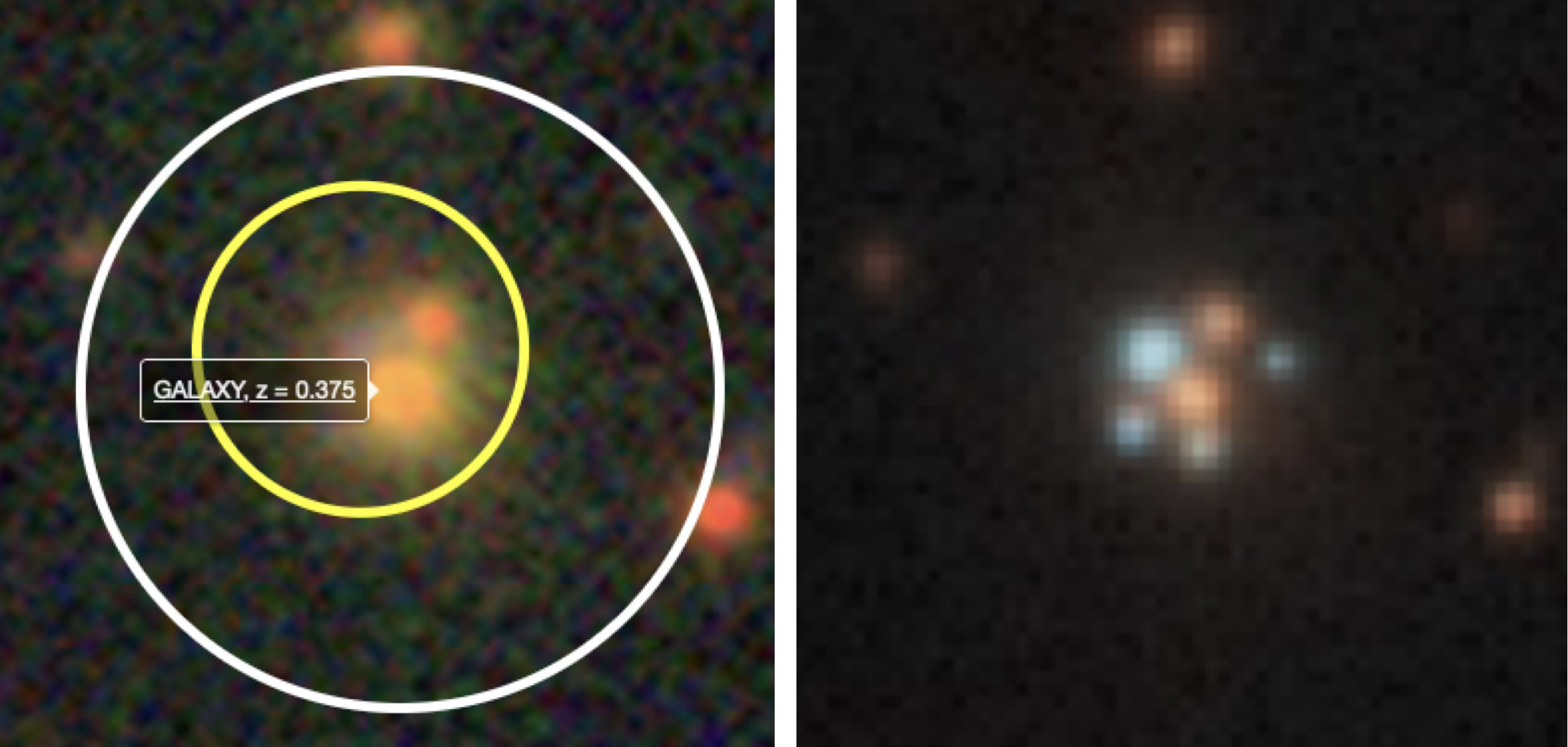} 
    \caption{
        \textbf{Archival and follow-up imaging of the lensed SN~2025wny.}
        (Left) The pre-explosion archival image from the DESI Legacy Surveys (LS) DR10, showing the host galaxy redshift from DESI DR1. 
        (Right) The optical color composite image obtained by the Lulin One-meter Telescope (LOT) following the alert from DETECT. 
        The prompt acquisition of this follow-up image highlights the pipeline's capability to bridge archival survey data with rapid photometric verification for extreme transients.
    }
    \label{fig:desi_lot_winny}
\end{figure}

One illustrative example is SN~2025wny \citep{2025arXiv251021694T,2025ApJ...995L..17J}, the first gravitationally lensed SLSN. Its lens galaxy is listed as a candidate in HOLISMOKES~II \citep{2020A&A...644A.163C}, but in 2025 September the pipeline did not yet match against lens catalogues, the event was highlighted in the DETECT report of 2025 September 1 purely by luminosity, because, with the DESI galaxy at $z=0.375$ as host, its reported $r$-band magnitude of $\sim$19.1 corresponded to $M\approx-22.5$, more than three magnitudes brighter than a normal supernova. The reviewer then recognised the lens-like configuration in the Legacy Surveys image and follow-up was requested. This placed the source among the most luminous SN-like transients and motivated follow-up coordination (Figure~\ref{fig:desi_lot_winny}). We subsequently contributed a LOT confirmation image to \citet{2025arXiv251021694T}.

\subsubsection{AT~2025aagx: Follow-up Analysis of a Rejected Lensed-Transient Candidate}

AT~2025aagx was initially flagged by DETECT as a potential lensed transient because it is located close to a bright elliptical galaxy and an adjacent arc-like feature suggestive of a background source. Such a configuration naturally merits further inspection in our screening framework, as a luminous transient projected near a massive early-type galaxy can be a plausible strong-lensing candidate.

To assess this possibility, we performed forward modeling of the system using \texttt{lenstronomy} \citep{Birrer2018}. For the lens mass distribution, we adopted a singular isothermal ellipsoid (SIE), whose dimensionless surface mass density is

\begin{equation}
\kappa(x,y)=\frac{1}{2}\left(\frac{\theta_{\rm E}}{\sqrt{q x^2+y^2/q}}\right),
\end{equation}

where \(\theta_{\rm E}\) is the circularized Einstein radius and \(q\) is the minor-to-major axis ratio. For the lens-galaxy light, we found that a two-component elliptical S\'ersic model provides a substantially better description than a single-component model, and also yields more reasonable parameters for the putative background source light, which we model with an additional elliptical S\'ersic profile. The best-fit mass-model parameters are summarized in Table~\ref{tab:aagx_mass_model}. The best-fit SIE model yields an Einstein radius of \(\theta_{\rm E}=0.975\arcsec\).

\begin{table}[htbp]
\centering
\caption{
Best-fit parameters of the lens mass model for AT~2025aagx.
The ellipticity parameters $(e_1,e_2)$ are converted to axis ratio $q$ and position angle $\phi$.
}
\label{tab:aagx_mass_model}
\begin{tabular}{lccccc}
\hline \hline
Model & $\theta_{\rm E}$ ($''$) & $q$ & $\phi$ (deg) & $x_0$ ($''$) & $y_0$ ($''$) \\
\hline
SIE & 0.975 & 0.584 & 16.0 & 0.065 & -0.048 \\
\hline
\end{tabular}
\end{table}

\begin{figure}[htbp]
    \centering
    \includegraphics[width=\linewidth]{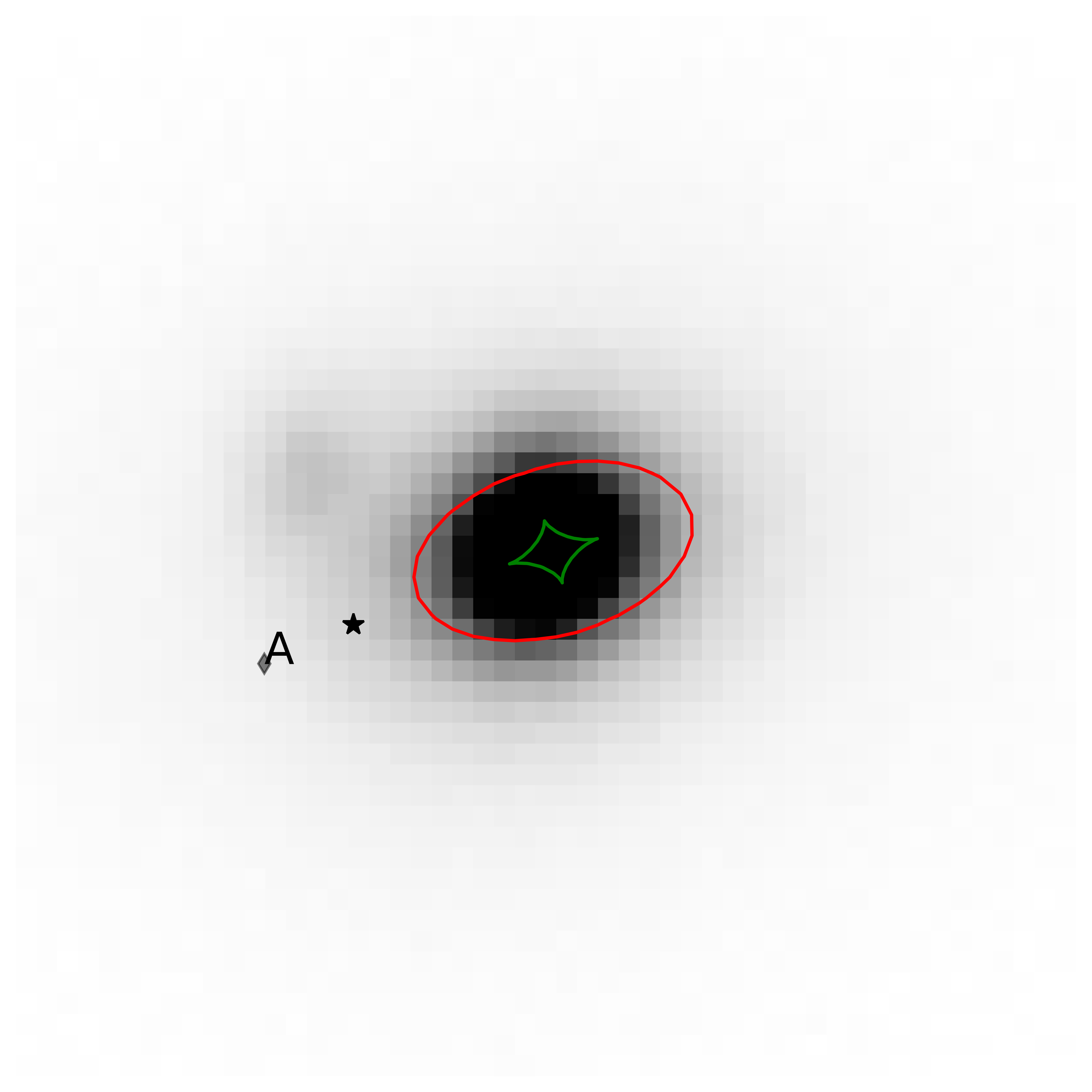}
    \caption{
        Best-fit SIE lens-model visualization for AT~2025aagx.
        The FITS cutout is shown as the background image, with the red and green curves denoting the critical curves and caustics, respectively.
        The position of AT~2025aagx is marked with a diamond.
        Treating this position as a putative lensed image, we map it back to the source plane and mark the inferred source position with a star.
        Under the best-fit model, no additional image is predicted, indicating that the current lensing configuration does not support AT~2025aagx as an observable multiply imaged transient.
    }
    \label{fig:aagx_lens_plot}
\end{figure}

To test the lensed-transient hypothesis, we treated the observed position of AT~2025aagx as one possible lensed image. Using the best-fit mass model, we mapped this image-plane position back to the source plane and then solved the lens equation for the inferred source position. Figure~\ref{fig:aagx_lens_plot} shows the resulting model geometry. In this reconstruction, the observed transient position can be reproduced as one image, but the model does not predict any additional detectable image. Thus, the inferred source position does not lead to a convincing observable multiple-image configuration.

We further subtracted the best-fit lens-galaxy light from the original image and inspected the residuals. No obvious counter image is seen in the residual map. Taken together, the imaging and lens-modeling results do not support AT~2025aagx as a convincing candidate for an observable strongly lensed transient.

\subsubsection{Additional Transients and Kilonova Search}

Beyond the detailed case studies of SN~2025wny and AT~2025aagx, the real-time DETECT pipeline successfully flagged several other anomalously bright transients during its 2025 operational run. By applying the photometric selection criteria, we identified multiple targets exhibiting exceptionally high inferred intrinsic luminosities. These targets automatically triggered alert notifications for rapid follow-up. A notable example is SN~2025sei, which was subsequently confirmed as an SLSN following robust host-galaxy association and follow-up photometry. The complete list of these high-probability SLSN candidates and other extreme transients identified in 2025 is provided in Appendix~\ref{sec:appendix_2025_slsn}.

Conversely, on the faint and fast-evolving end of the transient spectrum, we continuously monitored the incoming data stream for kilonova candidates (characterized by $M \gtrsim -17$\,mag and a rapid decline rate). However, we did not find any viable kilonova candidates from the real-time search during the 2025 operational period. This null detection is consistent with the extreme rarity and short observable window of these events, underscoring the necessity for the unprecedented depth and high-cadence observations that will be provided by upcoming facilities like the Rubin Observatory's LSST.

% =============================================================
% =============================================================
% =============================================================

\section{Discussions and Conclusions}
\label{sec:discussions}
\subsection{Comparison with Major Transient Surveys}
To assess the representativeness of our cross-matched sample, we compared the relative fractions of major SN types in our `Gold Sample' with those from three established surveys: the Lick Observatory SN Search (LOSS, \citealt{2011MNRAS.412.1441L}), the All-Sky Automated Survey for SNe (ASAS-SN, \citealt{2019MNRAS.484.1899H}), and the ZTF Bright Transient Survey (ZTF BTS, \citealt{2020ApJ...895...32F}). 

\begin{figure}[htbp]
    \centering
    \includegraphics[width=1\linewidth]{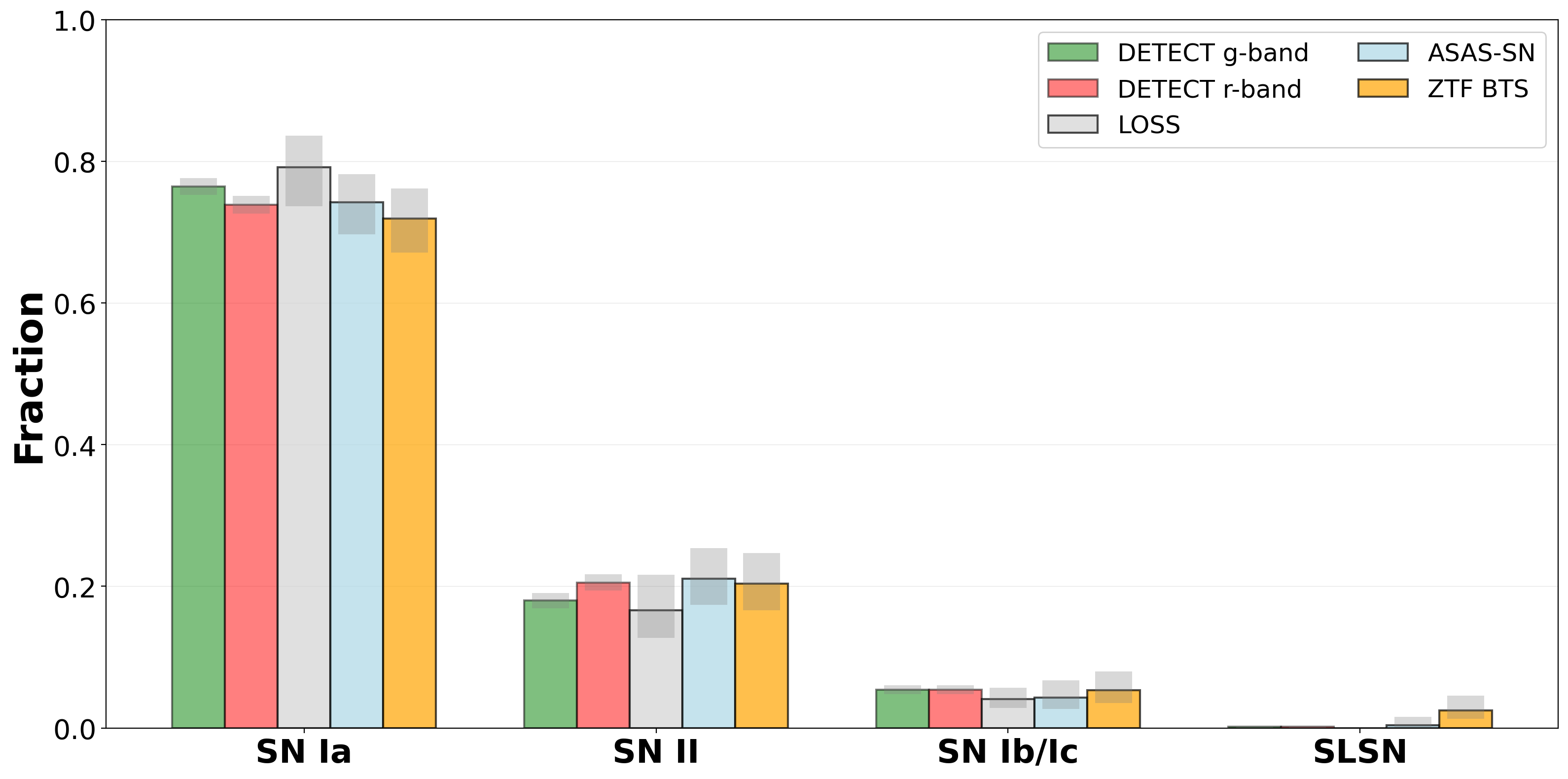}
    \caption{
        \textbf{Comparison of relative SN fractions between DETECT and major historical surveys.} 
        The bar chart displays the observed fractions of SN~Ia, SN~II, SN~Ib/Ic, and SLSN. 
        Green and Red bars represent the DETECT `Gold Sample' in $g$-band and $r$-band, respectively.
        Grey bars represent the volume-limited LOSS survey.
        Light blue bars represent the ASAS-SN sample.
        Orange bars represent the flux-limited ZTF Bright Transient Survey.
        The shaded regions at the top of each bar indicate the uncertainty (standard error for DETECT, published uncertainties for others). 
        The distribution of the DETECT sample closely mirrors that of ZTF and ASAS-SN, confirming that our cross-matching strategy yields a representative sample of the transient sky.
    }
    \label{fig:survey_comparison}
\end{figure}

The comparison is presented in Figure~\ref{fig:survey_comparison}. We calculated the observed fractions for the DETECT sample separately for the $g$- and $r$-bands to ensure consistency across filters. The results demonstrate that the population demographics of DETECT are largely consistent with other wide-field, flux-limited surveys (ZTF BTS and ASAS-SN). Specifically, SN~Ia constitutes the dominant class ($\sim 75\%$), followed by SN~II ($\sim 20\%$) and SN~Ib/c ($\sim 5\%$). 

This similarity confirms that our pipeline, despite relying on cross-matching with the specific footprint of DESI, does not introduce significant selection biases regarding SN types compared to untargeted all-sky surveys. The slight variations observed compared to LOSS can be attributed to the differences in survey strategies, as LOSS targeted specific nearby galaxies, whereas DETECT, ZTF, and ASAS-SN operate as wide-field surveys subject to flux limitations.

\subsection{Survey Depth and Detection Limits}
\begin{figure}[htbp]
    \centering
    \includegraphics[width=1\linewidth]{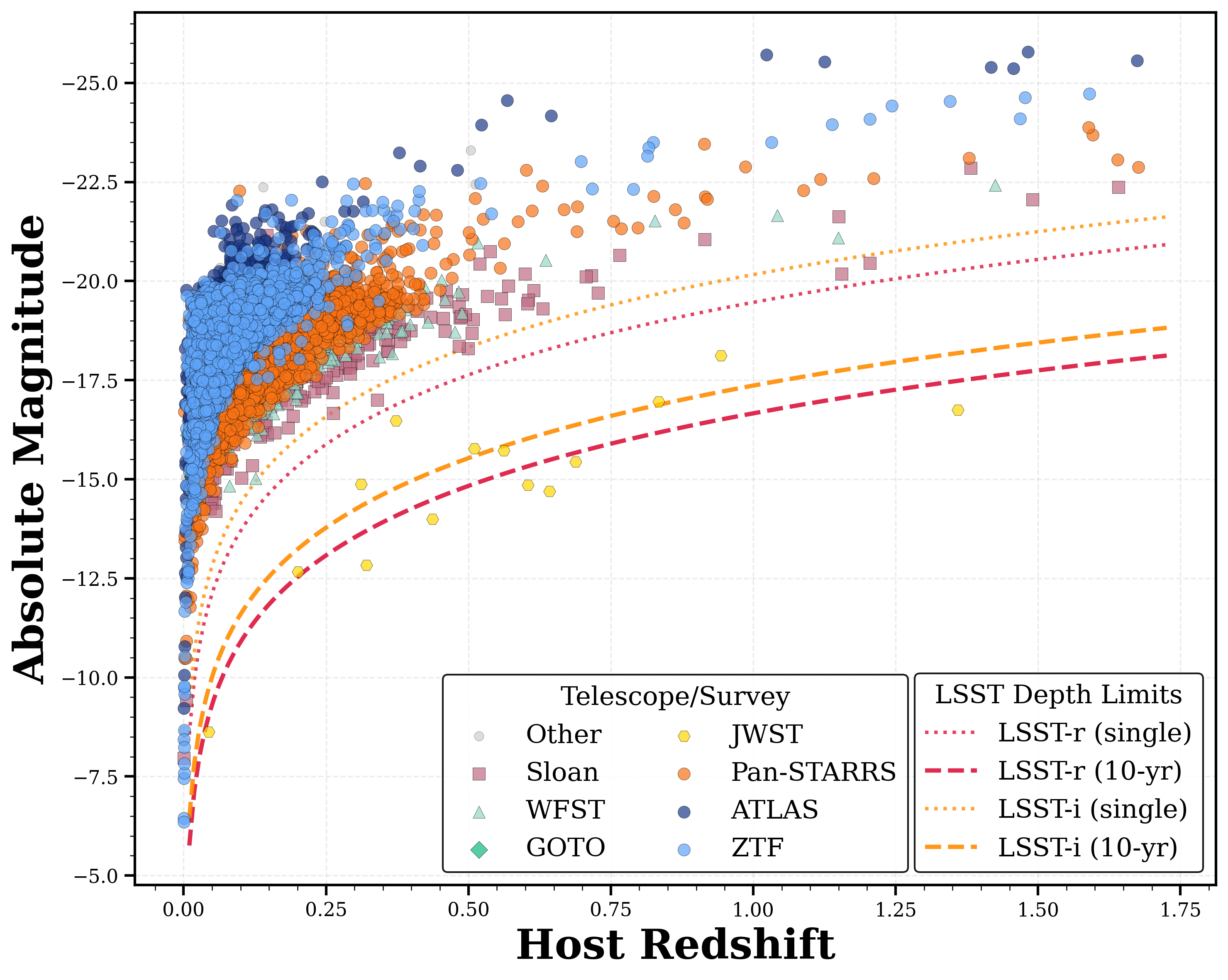}
    \caption{
    \textbf{Absolute magnitude versus redshift with LSST depth comparison.}
    Points show transients with spectroscopic host redshifts, colors, and shapes indicate the surveys providing the peak photometry (Pan-STARRS, ZTF, ATLAS, WFST, Sloan, GOTO, JWST, and others).
    Dashed curves denote LSST single-visit limits and solid curves the 10-yr combined limits (in $r$ and $i$ filters), converted to absolute magnitude as a function of redshift.
    The distribution shows the expected selection boundary: at higher redshift, only intrinsically brighter events are detectable.
    Note: To maximize sample inclusion for survey comparison, absolute magnitudes in this figure are derived from the brightest observed photometric point, rather than the fitted peak magnitudes used in Section~\ref{sec:abs_mag_analysis}.
    }
    \label{fig:M_z_survey_AT}
\end{figure}

To evaluate the detection capabilities of current time-domain surveys, Figure~\ref{fig:M_z_survey_AT} presents the relationship between absolute magnitude and redshift for our cross-matched sample. Crucially, for this specific survey-depth analysis, absolute magnitudes were derived using the ``brightest observed magnitude'' of each event. This methodology differs from the parameterized light-curve fitting approach.
While the fitting method yields higher physical precision for the ``Gold Sample,'' utilizing the brightest observed magnitude permits the inclusion of transients with sparse light curves (e.g., $N<5$) that would otherwise be excluded by strict quality cuts. This approach ensures a more complete sample for assessing the effective detection boundaries.

In Figure~\ref{fig:M_z_survey_AT}, points are color-coded by the telescope or survey providing the peak photometric data. The distribution reveals a clear observational bias: as redshift increases, only intrinsically brighter transients are detected. Specifically, at $z \approx 0.5$, targets generally require an absolute magnitude brighter than $M \approx -20$ to be observed. This trend directly reflects the limiting magnitudes of current instruments. Notably, deeper surveys such as Pan-STARRS, WFST, and SDSS account for the majority of detections at the faint end of the distribution. 

To contextualize future capabilities, we overlay the predicted single-visit and 10-year co-added limiting magnitudes for the Legacy Survey of Space and Time (LSST) \citep{2022ApJS..258....1B}. The comparison demonstrates that LSST will significantly deepen the detection horizon, likely becoming a game-changer for faint transient discovery.

\subsection{Future Prospects: Public Access and the LSST Era}
DETECT's output is served through the Kinder Project web interface, where the review queue, the derived quantities and the finders are available to the team, the same interface records the reviewers' decisions, which feed back into the pipeline. Opening the queue to the community, and matching against spectroscopic surveys beyond DESI, are the next steps.

Looking further ahead, DETECT serves as a strategic prototype for the upcoming Vera C. Rubin Observatory's Legacy Survey of Space and Time (LSST). As shown in Figure~\ref{fig:M_z_survey_AT}, LSST will push the detection horizon significantly deeper. The volume of alerts will increase by orders of magnitude \citep{2022ApJS..258....1B}, making purely photometric classification methods—like the absolute magnitude screening demonstrated here—indispensable. We plan to adapt the DETECT architecture to ingest the streams of the LSST community brokers \citep[ALeRCE, Fink, Lasair, ANTARES, AMPEL and others][]{2021AJ....161..242F, 2021MNRAS.501.3272M, 2019RNAAS...3...26S, 2021AJ....161..107M, 2019A&A...631A.147N}, ensuring that we can efficiently identify and prioritize high-value transients such as kilonovae and SLSNe in the petabyte-scale era.

% =============================================================
%  ACKNOWLEDGMENTS / SOFTWARE / FACILITIES
% =============================================================
\begin{acknowledgments}
% Person
We thank Stephen Smartt for comments on the manuscript and for discussions on the use of ATLAS and Pan-STARRS data, Ting-Wen Lan for guidance on the DESI data products, and Yen-Chen Pan for discussions on host-galaxy association. 

% Fellow
Y.-H. Lee, T.-W. Chen, A. Aryan, A. Sankar.K, K. N.-T. Ho, A. Dutta, M.-H. Lee, C.-H. Lai, C.-Y. Chen, T. Wang, P.-C. Hsu acknowledge the financial support from the Yushan Fellow Program by the Ministry of
Education, Taiwan (MOE-111-YSFMS-0008-001-P1) and the National Science and Technology Council,
Taiwan (NSTC grant 114-2112-M-008-021-MY3).

% Lulin
This publication has made use of data collected at Lulin Observatory, partly supported by the TAOvA with the NSTC grant 114-2740-M-008-002.

% Other
C-J.L. is supported by the NSTC grant 114-2112-M-004-001- MY2 from the National Science and Technology Council of Taiwan.

% DESI
This research used data obtained with the Dark Energy Spectroscopic Instrument (DESI). DESI construction and operations are managed by the Lawrence Berkeley National Laboratory. This material is based upon work supported by the U.S. Department of Energy, Office of Science, Office of High-Energy Physics, under Contract No. DE–AC02–05CH11231, and by the National Energy Research Scientific Computing Center, a DOE Office of Science User Facility under the same contract. Additional support for DESI was provided by the U.S. National Science Foundation (NSF), Division of Astronomical Sciences under Contract No. AST-0950945 to the NSF’s National Optical-Infrared Astronomy Research Laboratory, the Science and Technology Facilities Council of the United Kingdom, the Gordon and Betty Moore Foundation, the Heising-Simons Foundation, the French Alternative Energies and Atomic Energy Commission (CEA), the National Council of Humanities, Science and Technology of Mexico (CONAHCYT), the Ministry of Science and Innovation of Spain (MICINN), and by the DESI Member Institutions: www.desi.lbl.gov/collaborating-institutions. The DESI collaboration is honored to be permitted to conduct scientific research on I’oligam Du’ag (Kitt Peak), a mountain with particular significance to the Tohono O’odham Nation. Any opinions, findings, and conclusions or recommendations expressed in this material are those of the author(s) and do not necessarily reflect the views of the U.S. National Science Foundation, the U.S. Department of Energy, or any of the listed funding agencies.

% astro data lab
This research uses services or data provided by the Astro Data Lab, which is part of the Community Science and Data Center (CSDC) Program of NSF NOIRLab. NOIRLab is operated by the Association of Universities for Research in Astronomy (AURA), Inc. under a cooperative agreement with the U.S. National Science Foundation.

% legacy survey
The DESI Legacy Imaging Surveys consist of three individual and complementary projects: the Dark Energy Camera Legacy Survey (DECaLS), the Beijing-Arizona Sky Survey (BASS), and the Mayall z-band Legacy Survey (MzLS). DECaLS, BASS, and MzLS together include data obtained, respectively, at the Blanco telescope, Cerro Tololo Inter-American Observatory, NSF’s NOIRLab, the Bok telescope, Steward Observatory, University of Arizona, and the Mayall telescope, Kitt Peak National Observatory, NOIRLab. NOIRLab is operated by the Association of Universities for Research in Astronomy (AURA) under a cooperative agreement with the National Science Foundation. Pipeline processing and analyses of the data were supported by NOIRLab and the Lawrence Berkeley National Laboratory (LBNL). Legacy Surveys also uses data products from the Near-Earth Object Wide-field Infrared Survey Explorer (NEOWISE), a project of the Jet Propulsion Laboratory/California Institute of Technology, funded by the National Aeronautics and Space Administration. Legacy Surveys was supported by: the Director, Office of Science, Office of High Energy Physics of the U.S. Department of Energy, the National Energy Research Scientific Computing Center, a DOE Office of Science User Facility, the U.S. National Science Foundation, Division of Astronomical Sciences, the National Astronomical Observatories of China, the Chinese Academy of Sciences and the Chinese National Natural Science Foundation. LBNL is managed by the Regents of the University of California under contract to the U.S. Department of Energy. The complete acknowledgments can be found at https://www.legacysurvey.org/acknowledgment/.

% ATLAS
This work has made use of data from the Asteroid Terrestrial-impact Last Alert System (ATLAS) project. ATLAS is primarily funded to search for near earth asteroids through NASA grants NN12AR55G, 80NSSC18K0284, and 80NSSC18K1575, byproducts of the NEO search include images and catalogs from the survey area. The ATLAS science products have been made possible through the contributions of the University of Hawaii Institute for Astronomy, the Queen's University Belfast, the Space Telescope Science Institute, the South African Astronomical Observatory (SAAO), and the Millennium Institute of Astrophysics (MAS), Chile.

% PANSTARRS
The Pan-STARRS1 Surveys (PS1) and the PS1 public science archive have been made possible through contributions by the Institute for Astronomy, the University of Hawaii, the Pan-STARRS Project Office, the Max-Planck Society and its participating institutes, the Max Planck Institute for Astronomy, Heidelberg and the Max Planck Institute for Extraterrestrial Physics, Garching, The Johns Hopkins University, Durham University, the University of Edinburgh, the Queen's University Belfast, the Harvard-Smithsonian Center for Astrophysics, the Las Cumbres Observatory Global Telescope Network Incorporated, the National Central University of Taiwan, the Space Telescope Science Institute, the National Aeronautics and Space Administration under Grant No. NNX08AR22G issued through the Planetary Science Division of the NASA Science Mission Directorate, the National Science Foundation Grant No. AST–1238877, the University of Maryland, Eotvos Lorand University (ELTE), the Los Alamos National Laboratory, and the Gordon and Betty Moore Foundation.

% TNS
We acknowledge the use of the Transient Name Server (TNS) for the real-time ingestion of transient reports.

\vspace{1em}
The \textbf{DETECT} pipeline architecture and implementation were developed by Yu-Hsing Lee. The authors explicitly acknowledge the use of Large Language Models (including Claude and Google Gemini) and Grammarly for assistance with grammatical editing. The scientific content, analysis, and conclusions remain the sole responsibility of the authors.
\end{acknowledgments}

\software{
    \texttt{Python} \citep{1995python},
    \texttt{Astropy} \citep{2013A&A...558A..33A, 2018AJ....156..123A, 2022ApJ...935..167A},
    \texttt{Astroquery} \citep{2019AJ....157...98G},
    \texttt{Dustmaps} \citep{2018JOSS....3..695M},
    \texttt{Haffet} \citep{2021A&A...655A..90Y},
    \texttt{healpy} \citep{2005ApJ...622..759G, 2019JOSS....4.1298Z},
    \texttt{lenstronomy} \citep{Birrer2018},
    \texttt{POSSIS} \citep{2019MNRAS.489.5037B},
    \texttt{The Tractor} \citep{2016ascl.soft04008L},
    \texttt{Vizier} \citep{2000A&AS..143...23O},
}

%% ============================================================

\bibliography{citation}{}
\bibliographystyle{aasjournalv7.1}

% =============================================================
% =============================================================
%%  APPENDIX
\appendix
%% ============================================================

\section{Absolute Magnitude Distribution of the Full Cross-matched Sample}
\label{sec:appendix_mag_dist}

In this appendix, we present the distribution of peak absolute magnitudes for the entire catalog of transients successfully cross-matched with DESI galaxy spectra. Unlike the ``Gold Sample'' used for the detailed peak absolute-magnitude distribution analysis in Section~\ref{sec:results}, this dataset includes all events where a spectroscopic redshift could be associated with the transient, regardless of the light curve quality or the number of detection epochs.

\begin{figure}[htbp]
    \centering
    \includegraphics[width=1\linewidth]{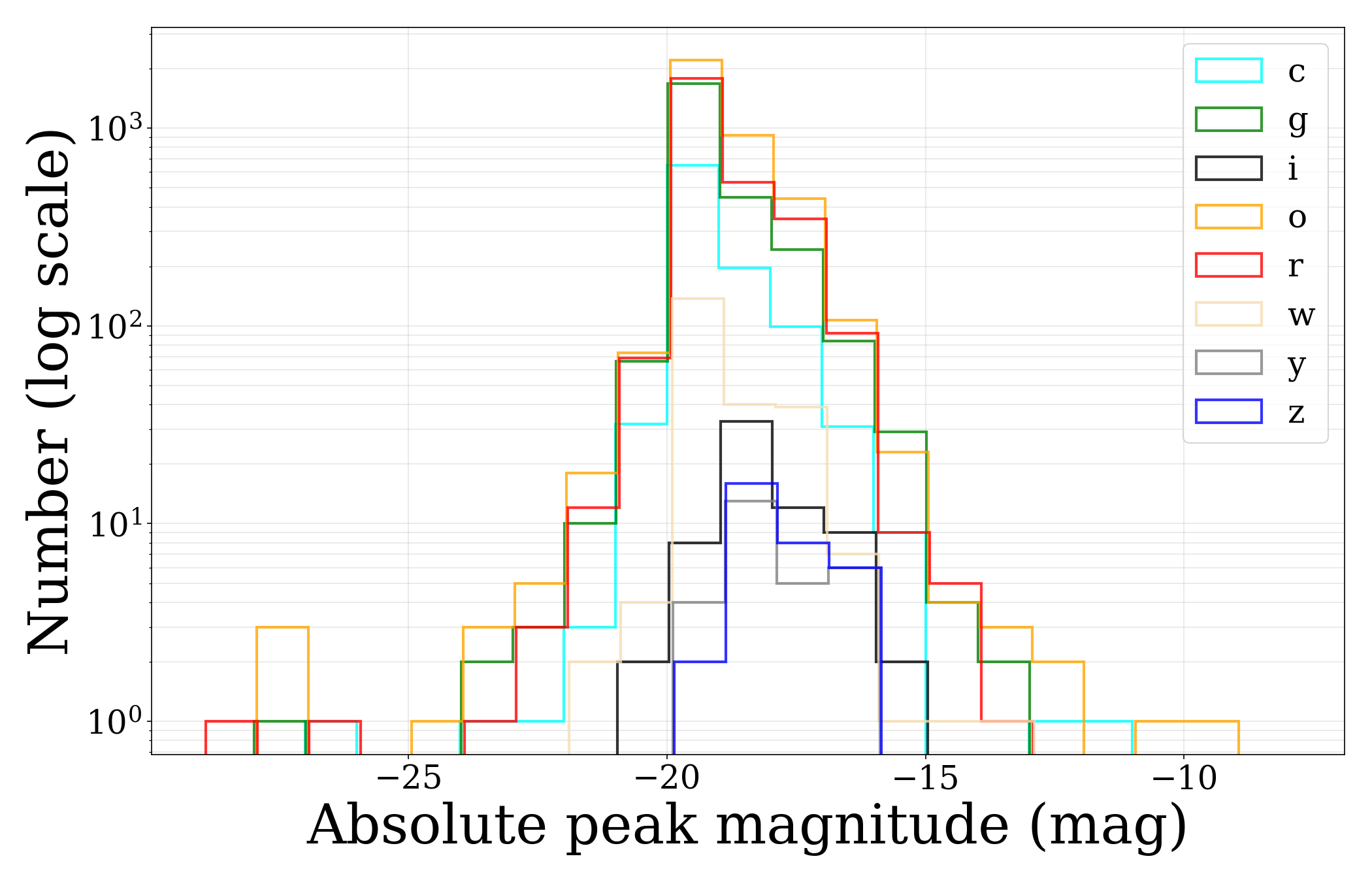}
    \caption{Distribution of absolute peak magnitudes for all cross-matched transients.
    The histograms are color-coded by filter.
    The y-axis represents the number of objects on a logarithmic scale.
    The distribution peaks near $M \approx -19.5$, typical of the SN~Ia population, but extends to both the high-luminosity ($M < -22$) and low-luminosity ($M > -15$) regimes.}
    \label{fig:all_at_mag_dist}
\end{figure}

Figure~\ref{fig:all_at_mag_dist} displays the histograms of the absolute peak magnitudes, separated by the photometric filter used for the observation. The distribution exhibits a prominent peak around $M \approx -19$ mag, which is consistent with the anticipated dominance of Type Ia SNe in flux-limited surveys. 

\section{List of High-Probability SLSN Candidates in archive}
\label{sec:appendix_candidates}

In this appendix we list the five unclassified transients that satisfy the high-contrast selection ($M_{r,\rm AT}<-21$ and $M_{r,\rm host}-M_{r,\rm AT}>0$) and are neither WISE AGN candidates nor unphysically bright (purple circles in Figure~\ref{fig:hm_plot_r_filter}). Host magnitudes are Legacy Surveys DR10 model magnitudes (Section~\ref{sec:host_transient}).

\begin{table}[htbp]
    \centering
    \caption{
        \textbf{Unclassified high-contrast transients.}
        Host redshift ($z$), fitted peak absolute magnitude of the transient ($M_t$), absolute magnitude of the host ($M_h$), their difference and the offset from the host centre.
    }
    \label{tab:slsn_candidates}
    \begin{tabular}{l c c c c c}
        \hline \hline
        \textbf{Object} & $z$ & $M_t$ (mag) & $M_h$ (mag) & $M_h - M_t$ & Offset (\arcsec) \\
        \hline
        AT 2020bcb  & 0.315 & $-22.16$ & $-21.46$ & 0.70 & 3.3 \\
        AT 2020mmc  & 0.180 & $-21.05$ & $-18.81$ & 2.24 & 0.7 \\
        AT 2020nqa  & 0.220 & $-21.10$ & $-21.09$ & 0.02 & 0.9 \\
        AT 2021yac  & 0.216 & $-21.15$ & $-21.08$ & 0.07 & 1.1 \\
        AT 2024abzf & 0.192 & $-21.08$ & $-21.03$ & 0.05 & 0.2 \\
        \hline
    \end{tabular}
\end{table}

\section{List of Kilonova Candidates}
\label{sec:appendix_kilonova}

In this appendix, we present the physical parameters of the fast-evolving transients flagged as potential kilonova candidates by DETECT. These targets were identified based on their consistency with theoretical kilonova models and the AT~2017gfo baseline.

\begin{table}[htbp]
    \centering
    \caption{
        The table lists the object name, spectroscopic redshift ($z$), and the derived peak absolute magnitude ($M$).
    }
    \label{tab:kilonova_candidates}
    \begin{tabular}{l c c l}
        \hline \hline
        \textbf{Object Name} & \textbf{Redshift ($z$)} & \textbf{Peak $M$ (mag)}  & \textbf{Note} \\
        \hline
        AT 2020bgp  & 0.029 & $-14.328$ & Flag out after Pan-STARRS detection\\
        AT 2020abut & 0.055 & $-17.368$ & Flag out after ZTF force photometry\\
        \hline
    \end{tabular}
\end{table}

\section{List of Potential Gravitationally Lensed Transients}
\label{sec:appendix_lens}

In this appendix, we present the five candidates identified through the spatial cross-matching of the 2020--2024 TNS sample with the strong lens galaxy catalogs.

\begin{table}[htbp]
    \centering
    \caption{
        \textbf{Potential Lensed SN Candidates identified by DETECT.}
        Columns listed are: (1) Transient Name, (2) The source catalog of the lens candidate, (3) Spectroscopic redshift of the background source galaxy ($z_{\rm src}$), (4) Angular separation between the transient and the lens center, (5) Quality metric from the source catalog.
    }
    \label{tab:lensed_candidates}
    \begin{tabular}{l c c c c}
        \hline \hline
        \textbf{Object Name} & \textbf{Lens Source} & \textbf{Source Redshift ($z$)} & \textbf{Sep. ($''$)} & \textbf{Grade} \\
        \hline
        AT 2020vbd  & Hsu+25  & 0.551 & 1.83 & C \\
        AT 2020afbc & Hsu+25  & 2.590 & 2.12 & B \\
        AT 2021jgl  & Hsu+25  & 2.567 & 1.74 & C \\
        AT 2022tnn  & Hsu+25  & 0.546 & 1.84 & C \\
        AT 2024fuc  & Hsu+25  & 0.805 & 0.95 & A \\
        \hline
    \end{tabular}
\end{table}

\newpage
\section{SLSN Candidates Identified in the 2025 Operational Run}
\label{sec:appendix_2025_slsn}

In this appendix, we present the list of Superluminous Supernova (SLSN) candidates and other extreme transients identified by the DETECT pipeline during its real-time operational run in 2025. The absolute magnitudes ($M$) reported in Table~\ref{tab:2025_slsn_candidates} represent the inferred magnitudes at the exact epoch when the targets were automatically flagged by our system. We note a special case for SN~2025sei: while its initial absolute magnitude was fainter than $-20$ mag upon discovery (as listed in the table), subsequent follow-up observations and robust host galaxy confirmation revealed its peak absolute magnitude to be brighter than the $-20$ mag threshold, securing its classification as an SLSN.

\begin{table}[htbp]
    \centering
    \caption{
        The columns denote the transient name, classification type (with ``Unclassified'' denoting targets lacking spectroscopic confirmation), spectroscopic redshift of the associated host galaxy, the inferred absolute magnitude at the time of the DETECT alert, and additional remarks.
    }
    \label{tab:2025_slsn_candidates}
    \begin{tabular}{l l c c l}
        \hline \hline
        \textbf{Name} & \textbf{Type} & \textbf{Redshift ($z$)} & \textbf{$M$ (mag)} & \textbf{Note} \\
        \hline
        SN 2025wny  & SLSN-I       & 0.375 & $-22.530$ & Lens SLSN \\
        AT 2025acxy & Unclassified & 0.590 & $-23.455$ & - \\
        AT 2025adtl & Unclassified & 0.846 & $-23.316$ & Photometry only one epoch \\
        AT 2025aagx & Unclassified & 0.281 & $-22.266$ & Possible Lens \\
        AT 2025ygm  & Unclassified & 0.261 & $-21.375$ & Photometry only one epoch \\
        AT 2025xid  & Unclassified & 0.234 & $-21.260$ & Photometry only one epoch \\
        SN 2025fco  & SN II        & 0.076 & $-20.762$ & Interacting galaxy \\
        SN 2025ktf  & SLSN-I       & 0.133 & $-20.480$ & Interacting galaxy \\
        SN 2025sei  & SLSN-I       & 0.095 & $-18.780$ & Reached $M < -20$ after follow-up \\
        \hline
    \end{tabular}
\end{table}

% End =============================================================

\end{document}